\documentclass[
    aps,
    prb,
    reprint,
    superscriptaddress,
    amsmath,
    amssymb
]{revtex4-2}

\usepackage{graphicx}
\usepackage{dcolumn}
\usepackage{bm}
\usepackage{physics}
\usepackage{booktabs}
\usepackage{hyperref}

\usepackage{xcolor}

\begin{document}

\title{Domain walls in $R3c$-structured NaNbO$_3$}

\author{Jack G. Nedell}
\thanks{Contact author: jn479@cornell.edu}
\affiliation{Department of Physics, Cornell University, Ithaca, New York 14853, USA}

\author{Craig J. Fennie}
\thanks{Deceased.}
\affiliation{Department of Applied and Engineering Physics, Cornell University, Ithaca, New York 14853, USA}

\date{\today}

\begin{abstract}
In this work, we undertake a systematic first-principles study of structural domain walls in the low-temperature rhombohedral $R3c$ phase of perovskite oxide NaNbO$_3$. Employing a group-theoretic classification, we introduce eight unique classes of domain walls in the $R3c$ structure, four of which have a domain wall orientation determined uniquely by the conditions of elastic compatibility and charge neutrality, and perform structural relaxations for each representative domain wall class. We find that one class of 109° domain walls has a notably low energy of 9 mJ m$^{-2}$, comparable to the low-energy ferroelectric walls in BaTiO$_3$. Finally, we find that this class of 109° walls is compatible with the crystal twinning observed in experiments on the $R3c$ phase, and propose an atomistic mechanism for the $R3c \leftrightarrow Pbcm$ phase transition. 

\end{abstract}

\maketitle

\section{Introduction}

The structural phase diagram of sodium niobate (NaNbO$_3$, or NNO) is famously pathological among the perovskite oxides. Characteristic of the numerous exotic phases observed in NaNbO$_3$ are long-period octahedral rotation patterns, persistent polar or antipolar off-centering of Nb$^{5+}$ B-site cations within their octahedral coordination environment, and competition or coexistence between different polymorphs nearly degenerate in energy, including antiferroelectric-ferroelectric (AFE - FE) competition. The complex set of structural phase transitions over temperature in NNO was first cast into a single progression by a seminal work, \textit{The seven phases of sodium niobate} by Megaw in 1974 \cite{megaw_seven_1974}, representing the culmination of multiple intensive crystallographic investigations by Megaw and contemporaries in the preceding decades \cite{darlington_low-temperature_1973,sakowski-cowley_structure_1969,ahtee_structures_1972,glazer_structure_1972,vousden_structure_1951}.

Decades later, the identification of desirable functional properties in NaNbO$_3$ has renewed interest in the material. NNO and related niobate solid solutions have in particular been identified as excellent lead-free candidates for applications requiring large piezoelectric response \cite{saito_lead-free_2004,zhu_ultrahigh_2025,wang_large_2022}, and for applications in capacitive energy storage \cite{qi_ultrahigh_2019,xie_superior_2026,jiang_boosting_2026,chen_phase_2022}. Careful engineering of the crystal structure is crucial in these and other application areas, with the above works employing such structural levers as engineered (nano)domains and boundaries \cite{qi_ultrahigh_2019,zhu_ultrahigh_2025} and modification of structural phase boundaries \cite{chen_phase_2022}.

Experimentally, NNO is observed to host structural phase coexistence \cite{darlington_low-temperature_1973,glazer_studies_1973,jiang_local_2013,htet_atomic_2022}, which may be sensitive to synthesis method \cite{johnston_polar_2010}, hysteresis in temperature-induced phase transitions \cite{darlington_low-temperature_1973,lefkowitz_high-temperature_1966,mishra_competing_2007,mishra_phase_2011}, highly disordered octahedral tilts \cite{kp_electron_2025}, an irreversible electric-field-induced $\text{AFE} \rightarrow \text{FE}$ phase transition \cite{ulinzheev_phase_1997,zhang_electric-field-induced_2020,zhang_revealing_2021}, and high sensitivity of phase selection to morphological factors such as strain or particle and grain size \cite{shiratori_polymorphism_2005,shiratori_temperature-induced_2007,koruza_grain-size-induced_2017}, as also demonstrated more recently in NNO membranes \cite{wang_room-temperature_2026}. These diverse experimental findings have accordingly motivated multiple first-principles studies on NNO, which have succeeded in illuminating some of these observed structural complexities. Density functional theory (DFT) calculations verify a delicate energetic competition between the low-temperature phases \cite{amisi_structural_2023,mishra_competing_2007}, and high sensitivity of phase ordering to the choice of density functional \cite{machado_relative_2011} as well as applied epitaxial strain \cite{patel_properties_2021}. First-principles work has also provided crystal-chemical intuition for the stability and persistence of the room-temperature AFE phase and the irreversible AFE $\rightarrow$ FE transition pathway \cite{hadaeghi_origin_2024,kim_explicating_2022}, and even reproduced a qualitative structural phase diagram of NNO by Monte Carlo studies using a classical lattice Hamiltonian \cite{yang_understanding_2018}. 

In these experimental crystallographic works, crystal twinning appears repeatedly in connection with the phenomena of structural phase coexistence and transitions \cite{darlington_low-temperature_1973,glazer_studies_1973,zhang_revealing_2021}. Despite this, explicit treatment of the crystallographic domain walls that lie between crystal twin domains does not appear in the first-principles literature on NaNbO$_3$. Arguably, first-principles and other theoretical treatments of structural domain walls in perovskites (that is, excluding magnetic walls) remain relatively limited in scope, with the bulk of such studies focusing on domain walls in the simple (BaTiO$_3$-type) ferroelectric structures \cite{padilla_first-principles_1996,meyer_ab_2002,li_first-principles_2014,zhang_first-principles_2023}. Works that do extend the study of structural domain walls to nonpolar structural degrees of freedom (such as octahedral rotations) have succeeded in providing valuable insights into the associated structural physics \cite{wang_bifeo_2013,lubk_first-principles_2009,kvasov_structure_2016,schranz_polarization_2020}.

In this work, we study the structural domain walls in the low-temperature $R3c$ phase of NaNbO$_3$. We employ a systematic symmetry-based classification of possible domain walls and perform first-principles structural relaxations for each wall class. A detailed analysis of the structural geometry and energetics associated with these domain walls identifies a particularly low-energy class of domain walls, which we find corresponds to observed crystal twinning in the $R3c$ phase. This motivates us to construct a candidate atomistic mechanism for the structural phase transition $Pbcm \leftrightarrow R3c$ between the room- and low-temperature phases. Our explicit treatment of domain walls provides access to aspects of NNO’s microstructural physics, particularly with regard to crystalline disorder and phase coexistence, that are not captured by first-principles treatments of bulk phases alone, while also enabling a symmetry-based classification not readily available from molecular dynamics or Monte Carlo approaches.

In Sec.~\ref{sec:bulkphase}, we introduce the relevant candidate bulk phases of NaNbO$_3$ in the room-to-low-temperature regime and discuss the relevant features of the established structural physics. Then, in Sec.~\ref{sec:lg_theory} we introduce the necessary group-theoretic background and techniques for the analysis and classification of structural domains and domain walls, and apply it to the $R3c$ structure of NaNbO$_3$. In Sec.~\ref{sec:methods} we detail the methods of our first-principles domain wall calculations and analysis. In Sec.~\ref{sec:DFT_results} we discuss the results of these first-principles calculations, including domain-wall energies, structural characteristics, and symmetry. Finally, in Sec.~\ref{sec:transition}, we connect the domain walls in $R3c$ with the structural phase transition between the low- and room-temperature phases $R3c$ and $Pbcm$, constructing an atomistic pathway for the transition consistent with the observed twinning in the $R3c$ phase.

\begin{figure*}[t!]
    \centering
    \includegraphics[width=0.9\textwidth]{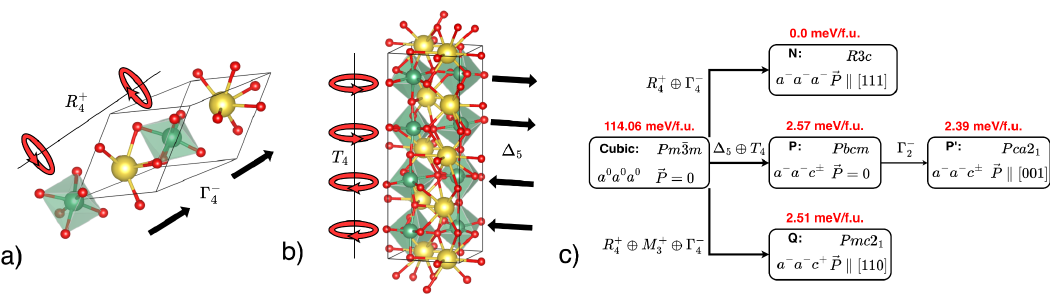}
    \caption{(a)-(b) two bulk phases of NaNbO$_3$. (a) shows the 10-atom primitive cell of low-temperature $R3c$, with the $R_4^+$ octahedral rotation mode and $\Gamma_4^-$ polar mode schematically diagrammed. (b) shows the 40-atom cell of the $Pbcm$ room-temperature AFE structure, with the ``$++--$'' type $T_4$ octahedral rotation mode and $\Delta_5$ antipolar mode diagrammed. (c) provides a group-subgroup tree for the cubic and four low-temperature bulk structures discussed. Each box in the tree (c) contains the label, space group, Glazer rotation pattern, as well as the DFT total energy with respect to the $R3c$ structure. Arrows in the tree are labeled with irreps of the parent that induce the symmetry-lowering from parent to child phase.} 
    \label{fig:bulk_and_group}
\end{figure*}

\section{Bulk phases of NNO}
\label{sec:bulkphase}

In this work, we focus on the low-to-room-temperature regime in NNO. In the most basic picture, this temperature range features an orthorhombic room-temperature phase with space group $Pbcm$, a rhombohedral low-temperature phase with space group $R3c$, and a hysteretic transition between them \cite{darlington_low-temperature_1973,mishra_phase_2011}. Here, we will also introduce the orthorhombic phases with $Pmc2_1$ and $Pca2_1$ symmetry, which may compete with the room-temperature $Pbcm$. Beyond these four phases, additional exotic and sometimes disordered polymorphs of pure NaNbO$_3$ have been observed experimentally in this temperature range \cite{shiratori_polymorphism_2005,shiratori_temperature-induced_2007,kp_electron_2025,ghanbari_strain-induced_2025}.

To introduce these four structures of NaNbO$_3$, we will discuss their geometric structural motifs, polarization, symmetry, and first-principles (Born-Oppenheimer) total energy. A summary of this information is diagrammed in Fig.~\ref{fig:bulk_and_group}, in the style of a group-subgroup tree. All computational methods involved are detailed later, in Sec.~\ref{sec:methods}. See Sec.~\ref{sec:lg_theory} for details on irrep modes and labeling convention. In this work, we primarily use pseudocubic crystallographic orientations, and this pseudocubic convention should be assumed unless otherwise stated.

\subsection{$R3c$}
The low-temperature phase of NNO, labeled the ``N'' phase in early crystallographic work, is polar with the $R3c$ space group. This structure is diagrammed in Fig.~\ref{fig:bulk_and_group}. This structure is identical to the $R3c$ room-temperature phase of LiNbO$_3$, often referred to as a ``LiNbO$_3$-like'' or ``pseudo-ilmenite'' structure, also found in BiFeO$_3$. Geometrically, this structure is related to the cubic perovskite parent by an $a^-a^-a^-$ octahedral rotation pattern, and a polar distortion pointing in the $[111]$ pseudocubic crystallographic direction, atomic displacement modes transforming respectively as the $R_4^+$ and $\Gamma_4^-$ irreps of space group $Pm\bar{3}m$. We find, in agreement with previous first-principles studies \cite{amisi_structural_2023}, that the $R3c$ structure is lowest in DFT total energy, by a comparable margin ($ \approx2.5  \text{ meV/f.u.}$) as previous works. The polarization in the $[111]_{\text{pc}}$ direction is calculated as $P = 53 \; \mu$C cm$^{-2}$, computed as $\Delta \mathbf{P} =\mathbf{P}_{R3c}-\mathbf{P}_{Pm\bar{3}m}$ by the Born effective charges and the atomic displacements away from the cubic parent. Notably, despite the identical structural degrees of freedom in the prototypical $R3c$ structure of LiNbO$_3$ (LNO) and $R3c$ NNO, atomic positions in NNO are closer to the reference cubic lattice. Comparing the rotation angle $\theta$ of the O$^{2-}$ octahedra about $[111]$, octahedra in NNO rotate about $12^\circ$, while in LNO this rotation angle is closer to $23^\circ$. The $R3c$ phase is observed in approximately the temperature range 0--160 K, but phase coexistence and hysteresis in the transition upon heating vs. cooling may significantly change the upper end of this window \cite{darlington_low-temperature_1973,jiang_local_2013,mishra_phase_2011}. Ref.~\cite{darlington_low-temperature_1973} also concluded that the low-temperature $R3c$ phase tends to form in equal fractions of two twin domains, related by a twin operation like $2_{001}$. This will serve as an important piece of evidence for our discussion in Sec.~\ref{sec:transition}.

\subsection{$Pbcm$}

The most commonly described room-temperature polymorph of NNO has space group $Pbcm$, and is labeled by Megaw as the ``P'' phase. This phase is generally referred to as the room-temperature antiferroelectric (AFE) structure; we will discuss its ferroelectric counterpart in the next subsection. This phase has a 40-atom primitive cell with dimensions $\sqrt{2} \times \sqrt{2} \times 4$ relative to the primitive cubic lattice, and features a complex (non-Glazer) octahedral rotation pattern, which we will denote $a^-a^-c^{\pm}$. This structure is diagrammed in Fig.~\ref{fig:bulk_and_group}. The atomic displacements that produce this pattern transform as the irreps $R_4^+$ and $T_4$ of $Pm\bar{3}m$. The $R_4^+$ mode gives the $a^-a^-$ symbols, denoting equal and out-of-phase octahedral rotations about $[100]$ and $[010]$; equivalently, an out-of-phase rotation about $[110]$. The $T_4$ mode, at $\mathbf{k}_T = \frac{2\pi}{a}(\tfrac{1}{2},\tfrac{1}{2},\tfrac{1}{4})$ results in a non-Glazer rotation pattern about $[001]$, corresponding to tilt angles with signs like $++--$ traveling along $[001]$. The other important atomic displacement mode transforms as irrep $\Delta_5$ at $\mathbf{k}_\Delta = \frac{2\pi}{a}(0,0,\tfrac{1}{4})$, which corresponds to antipolar displacements in the $[110]$ direction, switching sign in the same $++--$ pattern as the $T_4$ rotations while traveling along the long axis of the cell ($\mathbf{c} \parallel[001]$). The $Pbcm$ phase is found approximately in the range 160--633 K, making this the most relevant portion of the phase diagram for room-temperature functional applications. Crystallographic and first-principles studies of the $Pbcm$ structure may be found in Refs.~\cite{sakowski-cowley_structure_1969,jiang_local_2013,htet_atomic_2022,hadaeghi_origin_2024,amisi_structural_2023}. 

\subsection{$Pmc2_1$}

The additional metastable room-temperature polymorph of NNO, commonly labeled the ``Q'' phase, is the ferroelectric (FE) counterpart to the $Pbcm$ phase. We find the polarization of this phase to be $45 \; \mu$C cm$^{-2}$. The $Pmc2_1$ primitive cell has 20 atoms, with cell dimensions $\sqrt{2} \times \sqrt{2} \times 2$ relative to the cubic parent. $Pmc2_1$ is a polar subgroup of $Pnma$, and the structure is a polar variant of the prototypical CaTiO$_3$-like $Pnma$ perovskite structure. The largest atomic displacement modes which make up the $Pmc2_1$ structure transform as the irreps $R_4^+$, $M_3^+$, and $\Gamma_4^-$.  Interestingly, the stability of this structure thwarts a very common trend of dynamical stability against polar distortion in the $Pnma$ perovskites \cite{benedek_why_2013}. The persistence of this instability is likely owed to the strength of the coupling between occupied O 2p$^6$ and unoccupied Nb 4d$^0$ orbitals \cite{hadaeghi_origin_2024,cohen_origin_1992}.

The appearance of the $Pmc2_1$ phase in experiment varies, and frequently $Pbcm$- and $Pmc2_1$-structured regions are found to coexist, with this coexistence varying depending on synthesis method \cite{johnston_polar_2010,shakhovoy_ferroelectric_2012}. Accordingly, we find the two phases to lie within $0.1 \text{ meV/f.u.}$ of one another. Geometrically, the $Pbcm$ primitive cell is effectively composed of a pair of twinned $Pmc2_1$ primitive cells, where an $m_{110}$ mirror twin operation flips both the polarization orientation and the sign of the $M_3^+$ in-phase octahedral rotation mode; this picture of $Pbcm$ as composed of twinned $Pmc2_1$ units stacked along the $[001]$ axis is presented in other works as a geometric mechanism for coexistence of and transitions between these polymorphs \cite{htet_atomic_2022,kim_explicating_2022}. The electric-field-induced transition from the AFE to FE phase is traditionally reported to be irreversible. Similarly, incomplete transitions from AFE to FE leading to mixed phases are reported in the literature \cite{zhang_electric-field-induced_2020}. However, a recent work demonstrated that phase engineering of room-temperature NNO with the introduction of CaHfO$_3$ and AgNbO$_3$ succeeds in stabilizing a truly reversible AFE-FE transition and associated robust double-hysteresis loop \cite{luo_well-defined_2023}.

\subsection{$Pca2_1$}
\label{sec:Pca2_1}
Finally, we comment on the polar orthorhombic $Pca2_1$ phase. This structure is a (maximal) polar subgroup of $Pbcm$, reached by the mode transforming as the $\Gamma_2^-$ irrep of $Pbcm$. The resulting polarization points along the long axis of the cell, the $[001]$ direction, and we calculate it to be relatively small, $17 \; \mu$C cm$^{-2}$. This phase and the associated dynamical instability of the $Pbcm$ structure have only been reported in first-principles work \cite{yang_understanding_2018,patel_properties_2021,amisi_structural_2023}, and not, to our knowledge, reported experimentally. This lack of experimental validation may plausibly correspond to two scenarios: a) that the true order of room-temperature NNO does tend to be polar, but that the relatively small polar displacement away from $Pbcm$ associated with this $Pca2_1$ ordering is small enough to have escaped detection in crystallographic investigations, or b) that the zone-center dynamical instability of the $Pbcm$ structure is suppressed, whether thermally, by disorder related to crystal twinning/phase coexistence at room temperature, or even by zero-point quantum fluctuations. Given the extremely shallow depth of the well associated with the $\Gamma_2^-$ mode that takes $Pbcm \rightarrow Pca2_1$ (only $ 0.18  \text{ meV/f.u.}$ from our calculations), we argue that either possibility remains viable. The classical effective Hamiltonian employed by Yang et al. \cite{yang_understanding_2018} notably does produce a $Pca2_1$ room-temperature ground state structure. We will demonstrate later that the $Pca2_1$ structure is a useful midpoint in constructing the mechanism for transition $Pbcm \leftrightarrow R3c$, but elucidating the sensitive competition between $Pca2_1$ and $Pbcm$ beyond the 0 K Born-Oppenheimer energy landscape is beyond the scope of this work.

\section{Domain wall group theory}
\label{sec:lg_theory}

Crystallographic \textit{domains} arise from the discussion of some low-symmetry ``child'' structure which is obtained from a higher-symmetry ``parent'' structure by some (series of) displacive phase transitions. Say the child structure with space group $F$ is related to the parent structure with space group $G$ by the real-space pattern of atomic displacements $\mathbf{u}_\kappa(\mathbf{R}_i)$, the displacement vector for atom $\kappa$ in the parent unit cell at $\mathbf{R}_i$. Suppose further that the child space group $F$ is the stabilizer (or isotropy) subgroup of the atomic displacement pattern under the action of the parent group $G$, i.e.,

\begin{equation}
    F = \text{Stab}_G(\{\mathbf{u}_\kappa(\mathbf{R}_i)\})
\end{equation} 

For such displacive transitions, it is further possible to decompose the symmetry-lowering displacement pattern $\mathbf{u}_\kappa(\mathbf{R}_i)$ that relates the child to the parent structure into a basis of modes corresponding to the irreducible representations of the parent space group, here $G = Pm\bar{3}m$. We follow the work of Hatch and Stokes \cite{stokes_isotropy_1988,hatch_complete_2001}, and their ISOTROPY suite of tools \cite{stokes_isotropy_nodate,campbell_isodisplace_2006,hatch_invariants_2003} for displacive phase transitions, alongside the Bilbao Crystallographic Server \cite{aroyo_bilbao_I_2006,aroyo_bilbao_II_2006}, are used in support of crystallographic group theory calculations in this work. The irreps of $Pm\bar{3}m$ are computed with the B-site at the cubic cell origin, and irreps are labeled according to the scheme of Miller and Love. 

This decomposition generally permits the identification of a set of primary order parameters (OPs) \cite{hatch_complete_2001}, that is, a minimal set of vectors, each lying in the carrier space of an irrep of $G$, such that the corresponding atomic displacement pattern achieves the correct symmetry-lowering. From here, we will collect these primary order parameters into a single composite order parameter $\phi$, which transforms under the direct sum of their respective irreps.

So, the displacement sending $G \rightarrow F <G$ is described by the order parameter $\phi$, and $\phi$ is stabilized by $F$, 

\begin{equation}
    F = \{ g \in G | g\phi =\phi\} =\text{Stab}_G(\phi)
\end{equation}

Consider the action of $g' \in G $ where $g' \notin F$ on the order parameter $\phi$. The resulting $\phi' = g'\phi \neq \phi$ has a conjugate stabilizer  
\begin{equation}
\label{eq:stab_conj}
    F' = \text{Stab}_G(\phi') = g'Fg'^{-1}
\end{equation}
which is isomorphic to $F$, but generally $F \neq F'$. The invariance of the parent Hamiltonian under $G$ identifies the structures with order parameters $\phi$ and $\phi'$ as symmetry-equivalent. These structures, differing only in their geometric relationship to the shared parent structure, are \textit{domains} or \textit{domain states} of the child structure in the parent.

By Eq.~(\ref{eq:stab_conj}), the set of order parameters corresponding to all of the unique domain states of the low-symmetry structure can be obtained as the orbit of any initial order parameter, which we will label $\phi_0$,

\begin{equation}
    \{\text{domains} \} \equiv \{\phi_i\} = \text{Orb}_G(\phi_0)
\end{equation}

The orbit-stabilizer theorem gives a bijection between this set of domains and the (left) coset decomposition $G / F$ of the parent $G$ in terms of any stabilizer, say $F$:

\begin{equation}
\label{eq:cosets}
    G = \bigcup_{i=0}^{n-1} h_iF
\end{equation}

where $n$ is the index $[G:F]$. Each of these cosets $h_iF$ consists exactly of the elements $g_i$ that send $\phi_0$ to $\phi_i$, that is 
$$g_i \phi_0 = \phi_i \; \forall g_i \in h_iF$$

We fix the coset representative $h_0 = e = (1|000)$. Note that since $F$ need not generally be a normal subgroup of $G$, the left cosets above need not correspond to a quotient group.

\subsection{Layer group theory of DWs}

A domain wall (DW) is, most generally, some extended two-dimensional manifold that is the boundary between two domains. In the special case of a planar boundary between domains, two dimensions of translational symmetry and three-dimensional point operations may be preserved. Thus the symmetry group of a planar DW is a layer group. This lends some useful symmetry-based machinery to the study of DWs.

The complete layer group theory of DWs was presented by Schranz et al. for domain walls in SrTiO$_3$ \cite{schranz_polarization_2020}. Here we will briefly introduce their method.

Given, say, a set of order parameters $\{ \phi_i \}$ labeling corresponding unique domain states, a planar DW is defined for the purposes of symmetry classification by:

\begin{equation}
    ( \phi_i | \hat{\mathbf{n}}, p|\phi_j)
\end{equation}

That is: an ordered pair of half-space domains $(\phi_i,\phi_j)$, the orientation of the planar boundary between them $\hat{\mathbf{n}}$, and the position $p$ specifying the wall equation $\hat{\mathbf{n}} \cdot \mathbf{r} = p$. In this work, we will use the convention that normal $\hat{\mathbf{n}}$ orients the wall by pointing from the first domain in the pair towards the second, such that 
$$( \phi_i | \hat{\mathbf{n}}, p|\phi_j) = ( \phi_j | -\hat{\mathbf{n}}, -p|\phi_i)$$

The procedure for finding the layer group of this DW can be summarized tersely as follows, in a condensed notation. According to the domain-exchanging identity above, a symmetry of the DW must either map each domain onto itself while preserving the wall position and orientation, or exactly exchange the domains with one another while preserving wall position, but inverting the wall orientation $\hat{\mathbf{n}}$. We will write this as:

\begin{equation}
    \label{eq:full_LG}
    G_\text{DW} = F_{ij} \cup \hat{F}_{ij}
\end{equation}

where $F_{ij}$ and $\hat{F}_{ij}$ are respectively the sets of domain-preserving and domain-exchanging symmetry operations. Written more explicitly, we define these sets of symmetry operations as:

\begin{equation}
    F_{ij} = \{ g \in G_i\cap G_j \; | \; g \cdot (\hat{\mathbf{n}},p) = (\hat{\mathbf{n}},p) \}
\end{equation}

\begin{equation}
\label{eq:domain_exchange}
    \hat{F}_{ij} = \{ g \in G_{ij}\cap G_{ji} \; | \; g \cdot (\hat{\mathbf{n}},p) = (-\hat{\mathbf{n}},-p) \}
\end{equation}

Here, $G_i = \text{Stab}_G(\phi_i) = h_i F h_i^{-1}$ are the stabilizer subgroups for each $\phi_i$, and $G_{ij} = h_j F h_i^{-1}$ are the sets (note, not groups!) of symmetry operations that send one domain into the other, $g\phi_i = \phi_j \; \forall \; g\in G_{ij}$. We undertook the practical computation of DW layer groups using this scheme, where the action of some symmetry on a given domain state can be calculated simply in the order parameter vector space.

\subsection{Enumerating the unique domain walls}
\label{section:enum}

Certain domain walls are equivalent by symmetry. Studies of crystallographic domain walls often use this implicitly, e.g., classifying ferroelectric walls between domains of $P4mm$-structured ferroelectric BaTiO$_3$ as 90° or 180°. This equivalence is easy to reason through geometrically for simple ferroelectrics. In our case of a composite order parameter, especially an order parameter with transformation properties differing from that of a polar vector, a formal treatment of said equivalence is valuable. Here we provide a brief formal commentary on this topic, geared towards the practical classification of domain walls. The group-theoretic treatment of domain-pair equivalence dates at least to the 1972 work of Janovec \cite{janovec_group_1972}, and a more intensive discussion of the following can be found therein.

Consider a generic DW, $W^{ij}_{\hat{\mathbf{n}},p} =( \phi_i | \hat{\mathbf{n}}, p|\phi_j)$. Recalling earlier arguments, we consider every wall related to this one by application of an operation $g$ belonging to the parent space group $G$ to be physically equivalent,

\begin{equation}
    W^{ij}_{\hat{\mathbf{n}},p} \sim g W^{ij}_{\hat{\mathbf{n}},p} \in \text{Orb}_G(W^{ij}_{\hat{\mathbf{n}},p})
\end{equation}

Thus, the DW orbits under $G$ define equivalence classes in the space of all walls. For any transformation of the ordered domain pair $(\phi_i,\phi_j)$, the wall normal and position will transform in an equivariant manner, so some choice $(\hat{\mathbf{n}},p)$ for one wall fixes an equivalent pair $(\hat{\mathbf{n}}_i,p_i)$ for each symmetry-related domain pair. Recall also the earlier identity:

\begin{equation}
\label{eq:wallid}
    W^{ij}_{\hat{\mathbf{n}},p} = W^{ji}_{-\hat{\mathbf{n}},-p}
\end{equation}

Treating the wall normal and position as free parameters, under Eq.~(\ref{eq:wallid}) we reduce the problem to the orbit of the \textit{unordered} pairs of domains, $\{\phi_i,\phi_j\}$. This result alone is sufficient for the computational classification of unique DWs. The set of all possible unordered domain-pairs is the disjoint union of a finite number of distinct domain-pair orbits. The set of unordered domain-pair orbits therefore provides an initial classification of the symmetry-distinct DWs. Choosing one representative domain pair from each orbit reduces the remaining classification problem to the possible wall geometries $(\hat{\mathbf n},p)$ associated with that pair, modulo the action of its stabilizer. 

\subsection{Classifying domain walls in $R3c$}

The $R3c$ structure of NaNbO$_3$ is related to the $Pm\bar{3}m$ cubic parent by two types of atomic displacement mode, an octahedral rotation mode and a polar displacement, transforming respectively as the $R_4^+$ and $\Gamma_4^-$ irreps of $Pm\bar{3}m$. We will work with 6-component composite order parameters in the representation-space $R_4^+ \oplus \Gamma_4^-$; these two concatenated order parameters have the transformation properties of a pseudovector and a polar vector. The matrix representatives for these irreps are constructed such that the OP basis components align with their associated pseudocubic Cartesian axes. A full table of the matrix representatives and a discussion thereof is included in Appendix~\ref{appendix:repres}.

In the carrier space of this representation, we write the order parameter for a reference $R3c$ structure as $\phi_0 = (\phi,\phi,\phi,P,P,P)$, indicating out-of-phase octahedral rotations about $[111]$, and polarization pointing along this rotation axis. Among the domain states of $R3c$, the values of scalars $\phi$ and $P$ stay fixed, with only the relative sign of each component varying, so we introduce a compact notation for the order parameter,

$$(\phi,\phi,\phi,P,P,P) \equiv [111]_+$$

Here, bracketed $[111]$ indicates the orientation of the octahedral rotation pseudovector, and the subscript $+$ indicates that the polarization points along this axis in the positive direction. Taking the orbit of this order parameter under $G =Pm\bar{3}m$, there are 16 possible domain states, which can be written generally as the 16 permutations $[\pm1\pm1\pm1]_\pm$. With the domain-pair orbit method described in Sec.~\ref{section:enum}, we find there to be eight unique classes of unordered domain pairs, including one trivial pair. 

We can restrict the space of DW orientations $\hat{\mathbf{n}}$ by a) enforcing elastic compatibility between domain states based on the corresponding pair of bulk strain states, and b) charge neutrality of the DW. Respectively, these conditions are, for some domains $\{\phi_1,\phi_2\}$ with corresponding infinitesimal strain tensors $\varepsilon_{1,2}$ and polarization vectors $\mathbf{P}_{1,2}$, as follows:

\begin{equation}
\label{eq:elast}
    (\varepsilon_1-\varepsilon_2)_{ij}dx_idx_j=0, \; d\vec{x} \perp \hat{\mathbf{n}}
\end{equation}

This is the elastic compatibility condition of Ref.~\cite{fousek_orientation_1969}; charge neutrality further requires

\begin{equation}
\label{eq:qarge}
    (\mathbf{P}_1-\mathbf{P}_2)\cdot \hat{\mathbf{n}} = 0
\end{equation}

that the bound charge of the DW vanishes. If the two domains have different polarization states, this condition constrains the normal $\hat{\mathbf{n}}$ to lie in a plane.

In Table~\ref{table:unique_walls}, we list the unique representative domain pairs and the restrictions on DW orientation between them, as determined by the above conditions. We also provide a label for the wall character, consisting of the angle made between the polarization vectors of the neighboring domains, and whether the change in octahedral rotation pattern is ``easy'' or ``hard'' for the charge-and-strain permitted DW orientation, that is, whether the atomic displacements due to the different octahedral rotation patterns on either side of the wall are continuous or discontinuous across the DW.

In their work on BiFeO$_3$ \cite{lubk_first-principles_2009}, Lubk and co-workers established a set of representative DWs for the same type of perovskite $R3c$ structure, working backwards from the same charge and strain compatibility conditions to find plausible walls, using the geometric character of such walls to produce a partial classification. The four 71° and 109° walls in our group-theoretic classification map exactly onto the 71°/109° ``continuous''/``discontinuous'' (easy/hard) walls presented in Ref.~\cite{lubk_first-principles_2009}, which correspond to domain pairs with a single wall orientation allowed by Eqs.~(\ref{eq:elast})--(\ref{eq:qarge}). 

We also recover the two classes of 180° walls suggested by Lubk, while further establishing that each admits a continuum of allowed orientations. Since the paired domains share a strain state, these walls are constrained only to have normals perpendicular to their common polarization axis. For consistency, we consider the same fixed normal $[0\bar{1}1]$ used by Lubk.

Our classification finally introduces the additional ``0° nontrivial'' wall, not mentioned by Lubk, a class of DWs between domains with identical polarization states but opposite-sign octahedral rotations. These could be called translational antiphase boundaries. Charge and strain impose no constraint on the wall normal, but for calculations we choose to consider the single $[111]$ wall orientation which renders this an ``easy'' wall.

\begin{table}[t]

\centering
\caption{Representative DWs in the perovskite $R3c$ structure. Each DW class is represented with a domain pair $\{ \phi_0,\phi_i \}$, with $\phi_i$ provided in the leftmost column.}
\label{table:unique_walls}
\begin{tabular}{lcccc}
\toprule
$\phi_i$ &
$\lvert \operatorname{Orb}_G\{\phi_0,\phi_i\}\rvert$ &
Wall-normal constraint &
Wall character \\
\midrule
$[111]_+$                    & 16 & --                              & $0^\circ$ trivial \\
$[\bar1\bar1\bar1]_-$        & 16 & --                              & $0^\circ$ nontrivial \\
$[111]_-$                    & 16 & $\hat{\mathbf n}\perp[111]$     & $180^\circ$ easy \\
$[\bar1\bar1\bar1]_+$        & 16 & $\hat{\mathbf n}\perp[111]$     & $180^\circ$ hard \\
$[\bar1 11]_+$               & 48 & $\hat{\mathbf n}\parallel[011]$ & $71^\circ$ hard \\
$[1\bar1\bar1]_-$            & 48 & $\hat{\mathbf n}\parallel[011]$ & $71^\circ$ easy \\
$[11\bar1]_-$                & 48 & $\hat{\mathbf n}\parallel[001]$ & $109^\circ$ easy \\
$[\bar1\bar1 1]_+$           & 48 & $\hat{\mathbf n}\parallel[001]$ & $109^\circ$ hard \\
\bottomrule
\end{tabular}
\end{table}

\section{Computational methods}
\label{sec:methods}

\subsection{First-principles calculations}

The first-principles calculations in this work were performed in the density-functional theory (DFT) package Quantum ESPRESSO, using the projector-augmented wave (PAW) method. We used
the Perdew-Burke-Ernzerhof functional as optimized for solids (PBEsol), obtained from the efficiency version of the SSSP library \cite{prandini_standard_2023}. PBEsol was previously identified as the best-performing functional for the structural features and energetics of NNO \cite{amisi_structural_2023}. The valence states for our
calculations are $2s^22p^63s^1$ for Na, $4s^24p^64d^45s^1$ for Nb and
$2s^22p^4$ for O. The Brillouin zone was sampled with an $8 \times 8 \times 8$ Monkhorst-Pack \textit{k}-point grid for the primitive cubic perovskite cell, and grids of equal or greater density for supercells thereof. With respect to the cubic lattice, the DW supercells had dimensions $\sqrt{2} \times \sqrt{2}  \times 16 $ (160 atoms) for the 109° walls, $2 \times \sqrt{2}  \times 8\sqrt{2} $ (160 atoms) for the 180° and 71° walls, and  $\sqrt{2} \times \sqrt{2}  \times 8\sqrt{3} $ (120 atoms) for the 0° wall.

The plane-wave energy cutoff was set to 75 Ry. Structural relaxations were performed by relaxing all atomic-position and lattice degrees of freedom. Atomic forces were converged to below 1e-4 Ry/bohr ($\approx$ 2.6 meV/Å) in the large DW supercells, and 1e-5 Ry/bohr in the smaller bulk-structure primitive cells. Cell relaxations were converged to residual pressure below 0.01 GPa.

\subsection{Domain wall computational approach}

DW supercell structures were initialized prior to relaxation beginning with a supercell of the cubic parent structure, constructed such that the $\mathbf{a}$ and $\mathbf{b}$ supercell lattice vectors are identically lattice vectors of the DW layer group and associated 2D discrete translations, and the $\mathbf{c}\parallel\hat{\mathbf{n}}$ supercell axis lies along the DW normal. The periodic DW superstructure is created from this cubic supercell by the application of sets of atomic displacements, interpolated between the two domains by a continuous step-like function $\Theta(z)$, periodic as a square wave in the $c$-axis coordinate $z$,

$$ \mathbf{u}_{\text{DW}} = \Theta(z)\mathbf{u}_{\phi_1} + (1-\Theta(z))\mathbf{u}_{\phi_2} $$

Construction of the supercell this way, containing two equal-thickness domains, may introduce new point symmetries to the supercell, not found in the layer group of an isolated DW. Furthermore, the symmetry and energetics of the walls may vary, depending on the initial position $p$ chosen for the DW center. 
In this work, we choose to initialize all DWs with $p=0$; this position corresponds to a parent-lattice plane containing the origin of the parent cubic cell, the midpoint between two equal-thickness domains, each containing an integer number of $R3c$ primitive units. We do not explore exhaustively the metastable position-dependent wall configurations for every DW class, but we do find that $p=0$ is the lowest-energy position for the 109° easy wall of most interest. Further discussion on DW position $p$ and supercell symmetries is provided in Appendix~\ref{appendix:supercell}.

The first-principles calculation of DW excess energy per unit area relative to the bulk (single-domain) crystal structure is generally nontrivial, as different mechanical boundary conditions define thermodynamically different excess quantities. Multiple distinct schemes have accordingly been implemented in the past \cite{lubk_first-principles_2009,zhang_first-principles_2023,meyer_ab_2002}. Our approach in this work is to compare the DFT total energy of bulk and DW structures under zero applied stress. Since our focus is on the sensitive structural energetics and geometry in the vicinity of the DWs, we do not wish to impose the constraint of zero rhombohedral strain, as chosen in \cite{lubk_first-principles_2009}. 

\begin{figure}[h]
    \centering
    \includegraphics[width=\columnwidth]{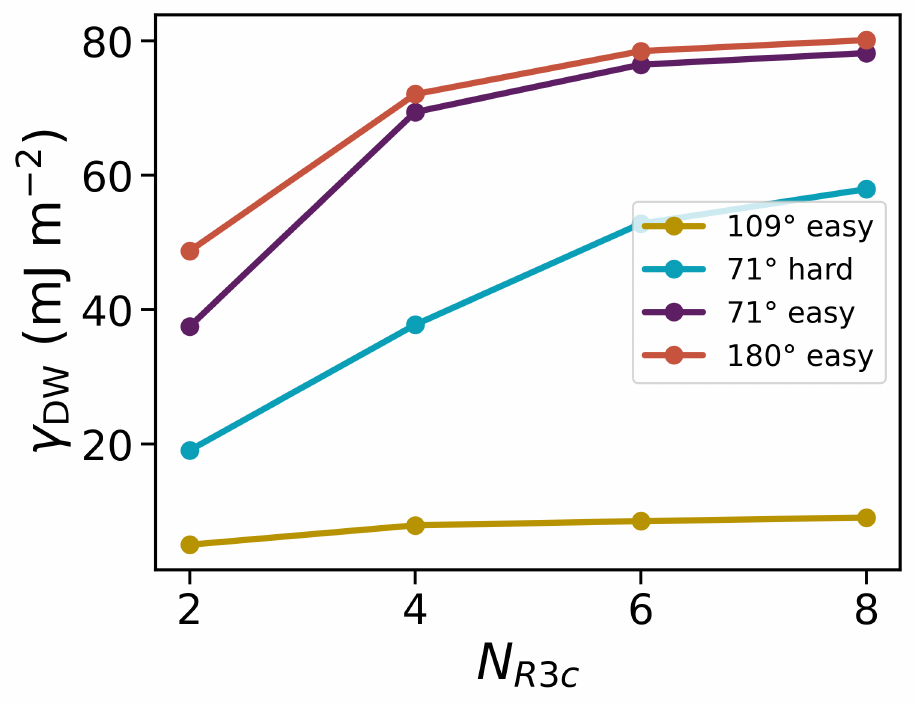}
    \caption{DW energy per unit area, $\gamma_{\text{DW}}$ as computed with Eq.~(\ref{eq:dw_gamma}) for increasing number $N_{R3c}$ of primitive 10-atom $R3c$ units in each domain.}
    \label{fig:convergence_E}
\end{figure}

We write the wall energy per unit area as:

\begin{equation}
    \label{eq:dw_gamma}
    \gamma_{\text{DW}}(N) \approx \frac{1}{2A}(E_{\text{DW}}^\text{super}(N) - E^0_{R3c}(N))
\end{equation}

where $A$ is the cross-sectional area of the relaxed DW supercell in the wall plane, and $N$ labels the domain thickness or number of $R3c$ primitive units in each domain. In Fig.~\ref{fig:convergence_E} we demonstrate the convergence of this expression towards a constant value for relaxed DW supercells with increasing domain thickness. The $N=8$ supercells show sufficient convergence, both in terms of DW energy and in their qualitative structural characteristics.

\begin{table*}[t]
\caption{Representative DWs, their layer groups, and energies in NaNbO$_3$ (NNO), with
corresponding BiFeO$_3$ (BFO) DW energies from Ref.~\cite{lubk_first-principles_2009}.
A dash indicates that no corresponding value is reported.}
\label{table:wall_energys}

\begin{tabular}{lcccc}
\toprule
& & &
\multicolumn{2}{c}{DW energies (mJ\,m$^{-2}$)} \\
\cmidrule(lr){4-5}
Representative wall
& Class label
& Layer group (No.)
& $\gamma_{N=8}$ (NNO)
& $\gamma_{\rm DW}$ (BFO)\cite{lubk_first-principles_2009} \\
\midrule

$\bigl([111]_+ \,\big|\, [0\bar{1}1],0 \,\big|\, [111]_-\bigr)$
& 180° easy & $p\bar{1}$ (2) & 80 & 829 \\

$\bigl([111]_+ \,\big|\, [0\bar{1}1],0 \,\big|\, [\bar{1}\bar{1}\bar{1}]_+\bigr)$
& 180° hard & $p\bar{1}$ (2) & --- & --- \\

$\bigl([111]_+ \,\big|\, [111],0 \,\big|\, [\bar{1}\bar{1}\bar{1}]_-\bigr)$
& 0° easy & $p3$ (65) & 173 & --- \\

$\bigl([111]_+ \,\big|\, [011],0 \,\big|\, [\bar{1}11]_+\bigr)$
& 71° hard & $pb11$ (12) & 58 & 436 \\

$\bigl([111]_+ \,\big|\, [011],0 \,\big|\, [1\bar{1}\bar{1}]_-\bigr)$
& 71° easy & $pb11$ (12) & 78 & 363 \\

$\bigl([111]_+ \,\big|\, [001],0 \,\big|\, [11\bar{1}]_-\bigr)$
& 109° easy & $pb11$ (12) & 9 & 205 \\

$\bigl([111]_+ \,\big|\, [001],0 \,\big|\, [\bar{1}\bar{1}1]_+\bigr)$
& 109° hard & $pb11$ (12) & 117 & 896 \\

\bottomrule
\end{tabular}
\end{table*}

\subsection{Structural analysis methods}
\label{sec:analysis_methods}

Typically, the most robust characterization tool for perovskite crystal structure is the method of symmetry-adapted modes. In this work, we have large and low-symmetry supercells, and do not wish to refer to the discrete translational symmetry along the long $\mathbf{c}$ axis. As such, we introduce local analogues of our order parameters. We will evaluate these local order parameters as they depend on the DW-normal coordinate, which we denote $\xi$. In plots, we will reference atomic-layer-derived order parameter values with the coordinate $\xi_0$ of the planar layer in the cubic parent cell, for simplicity. We introduce these briefly below; full order-parameter expressions and a description of their computational implementation are provided in Appendix~\ref{appendix:OP}.

The local ``polarization'' order parameter $\mathbf{P}(\xi_l)$ is defined in Eq.~(\ref{eq:pol_local}) and provides the Born effective charge (BEC) derived linear electric polarization, as computed for a region around the atomic layer indexed $l$ along the DW normal $\hat{\mathbf{n}}$. This should not be conflated with a true bulk polarization in the sense of the modern theory \cite{resta_theory_2007}, but serves as a suitable local $\Gamma_4^-$ type order parameter.

The local octahedral rotation order parameter $\bm{\phi}(\xi_l)$ is defined in Eq.~(\ref{eq:rot_local}), and is constructed using the numerically obtained rotation vector of each oxygen octahedron. Modulation of these rotation angles by the R point wavevector $\mathbf{k}_R = \frac{2\pi}{a}(\frac{1}{2},\frac{1}{2},\frac{1}{2})$ and averaging over each layer of octahedra yields the $R_4^+$-rotation-like local numerical value $\bm{\phi}(\xi_l)$ for the $l^\text{th}$ layer of octahedra along $\hat{\mathbf{n}}$.

We also seek a description of long-period components of the atomic displacement field that produce an effective position-dependent (inhomogeneous) contribution to the infinitesimal strain tensor, $\varepsilon_{ij}(\xi)$, that is we seek to identify whether smooth atomic displacements correctly compensate for the change in strain state between domains, a change which is not compatible with strain of the DW supercell. This is obtained by way of a Gaussian average over the atomic displacement field, and derivatives thereof. The expression for $\varepsilon_{ij}(\xi)$ is provided in Eq.~(\ref{eq:local_strain}).

\section{DW structure, symmetry, and energetics}
\label{sec:DFT_results}

Following the procedures outlined above, we computed the DW energy for each representative wall, presented in Table~\ref{table:wall_energys}. In this table, we also report the energies computed by Lubk et al. \cite{lubk_first-principles_2009} for equivalent walls in BiFeO$_3$ (BFO), where available. While hosting the same type of $R3c$ structure, BFO also hosts structure-coupled ferromagnetic ordering, and is generally ``more displaced'' relative to a $Pm\bar{3}m$ perovskite parent. Here we discuss the symmetric, geometric, and energetic features of the DW structures relaxed in our first-principles calculations. Note that the mechanical boundary conditions imposed for DW energy computation differ between this work and the BFO work of Ref.~\cite{lubk_first-principles_2009}, although this difference is unlikely to alter the qualitative energy-scale comparison.

\subsection{DW energetics and layer groups}

In Table~\ref{table:wall_energys} we report the DW energies, in units of mJ m$^{-2}$. In both NNO and BFO, the lowest-energy DWs belong to the 109° easy class. However, we find the DW energies in NNO are about one order of magnitude smaller than in BFO. The value of 9 mJ m$^{-2}$ we report for the 109° easy wall is comparable to the characteristically low-energy walls found in prototypical perovskite ferroelectric BaTiO$_3$ \cite{zhang_first-principles_2023,meyer_ab_2002}. This is a significant result of this work. We found that relaxing a metastable DW structure of the ``180° hard'' type was challenging, with long BFGS relaxation trajectories, very high energy compared to the bulk, and significant drift of the individual domains away from any bulk-like structure. As such, we omit this wall energy from the table above, and from further discussion. Note that the 0° easy octahedral-rotation-only DW introduced in this work is quite high in energy.

Table~\ref{table:wall_energys} also lists the DW layer groups (with their standard-setting labels), as computed with Eqs.~(\ref{eq:full_LG})--(\ref{eq:domain_exchange}). The 180° walls have $p\bar{1}$ symmetry, such that the only nontrivial point operation is $(\bar{1}|000)$. Said walls cannot move position $p$ without breaking this inversion. The 71° and 109° walls all have layer group $pb11$, \#12, with nontrivial symmetry $(m_{100}|\tfrac{1}{2}\tfrac{1}{2}0)$ as written in the setting of our DW supercell. This symmetry is not domain-exchanging and thus does not depend on the wall position $p$. The 0° easy wall has layer group $p3$ with a threefold rotational symmetry about the wall normal axis, written in the standard setting $(3^\pm_{001}|000)$; this also does not exchange the domains.

\begin{figure*}[th]
    \centering
    \includegraphics[width=\textwidth]{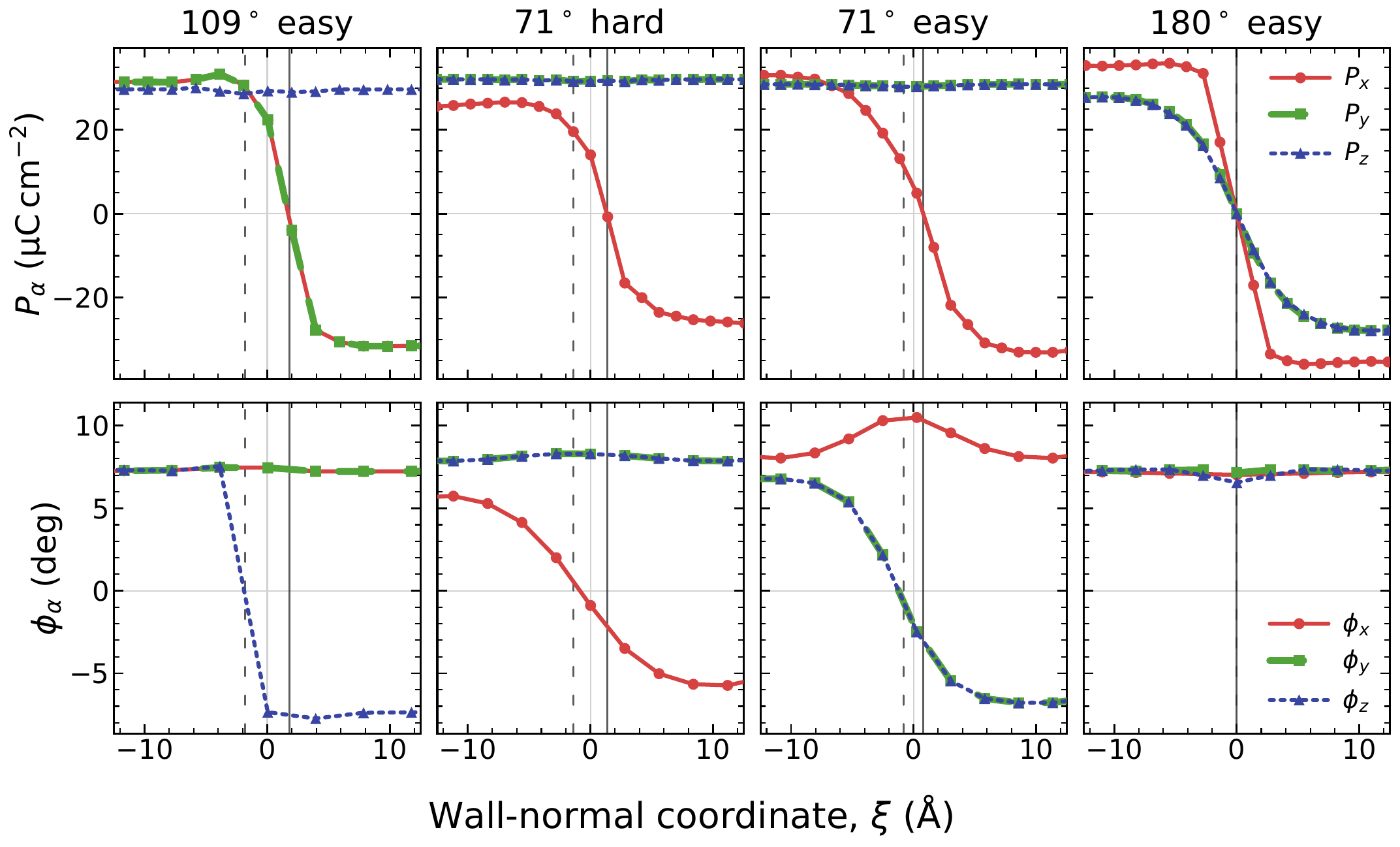}
    \caption{Local polarization (top row) and octahedral rotation (bottom row) order parameter profile, computed over a window of 25 Å around one of the two DWs in the DW supercell, for the 109° easy, 71° hard, 71° easy, and 180° easy DW representatives. Each point represents the value of the local order parameter averaged over a crystallographic plane or layer of octahedra along the DW-normal axis. The solid and dotted black vertical lines represent, respectively, the ``wall center'' for the individual polarization and octahedral rotation order parameters.}
    \label{fig:dw_local_order_parameters}
\end{figure*}

\begin{figure*}[t!]
    \centering
    \includegraphics[width=0.9\textwidth]{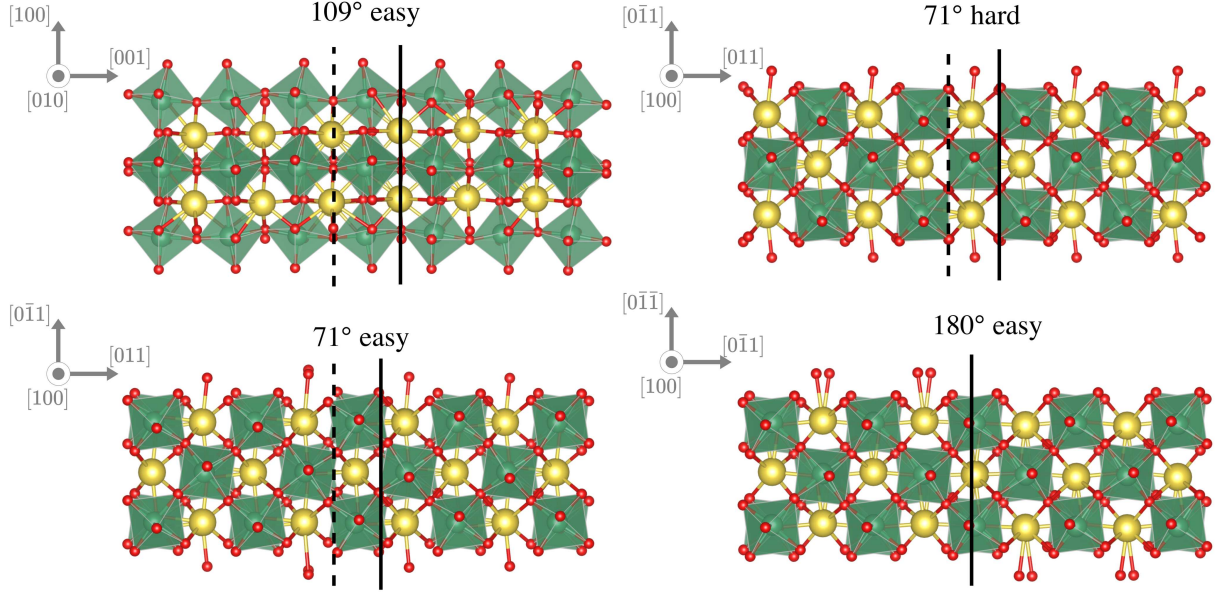}
    \caption{Segments of relaxed DW supercells for each of the four example walls. The atomic planes diagrammed here correspond approximately to the 25 Å window shown in Fig.~\ref{fig:dw_local_order_parameters}, and not to the entire 160-atom supercell. Each cell's orientation with respect to the pseudocubic crystallographic directions is indicated with a triad of axes to the upper left of the cell. The vertical solid and dashed black lines indicate respectively the center of the polarization and octahedral rotation wall profiles.}
    \label{fig:dw_vesta_diag}
\end{figure*}

\begin{figure}[t]
    \centering
    \includegraphics[width=\columnwidth]{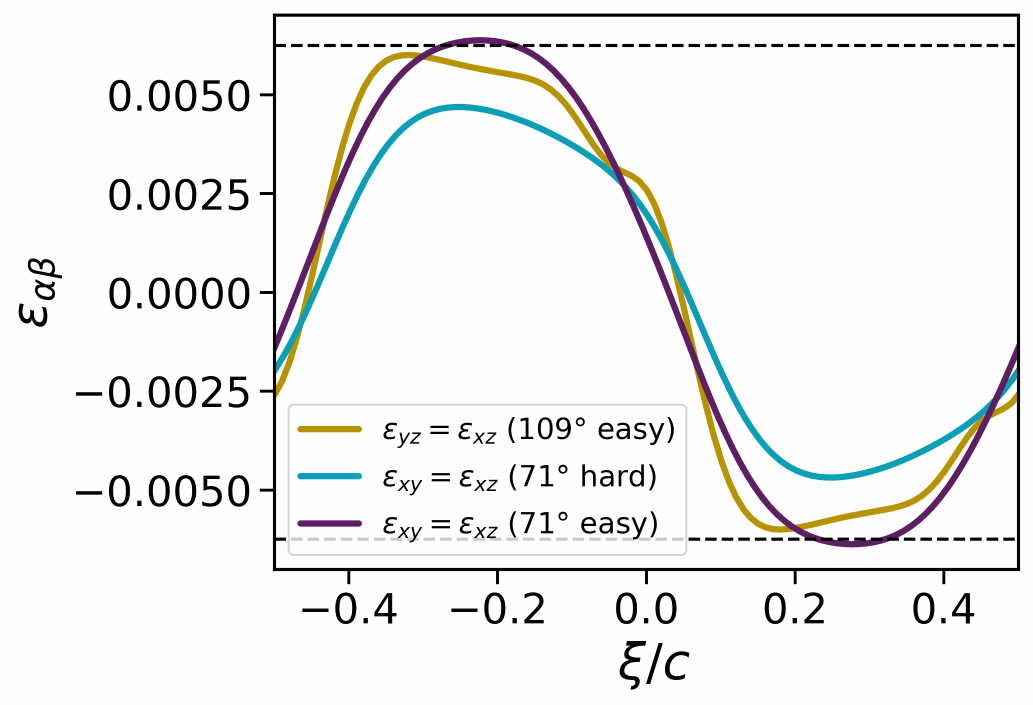}
    \caption{The local components of inhomogeneous strain in the relaxed 109° easy and 71° easy/hard DW supercells, derived from $Pm\bar{3}m \rightarrow \text{ DW}$ atomic displacement patterns with Eq.~(\ref{eq:local_strain}). Horizontal dotted lines indicate the magnitude of off-diagonal rhombohedral strain in relaxed bulk $R3c$ relative to $Pm\bar{3}m$. The x-axis $\xi/c$ indicates the fractional coordinate, relative to the wall center, along the DW-normal ($c$ axis) of the DW supercell.}
    \label{fig:dw_local_strain}
\end{figure}

\subsection{Structure of the relaxed DWs}
In Fig.~\ref{fig:dw_local_order_parameters} we plot the local polarization and octahedral-rotation order parameters, as defined in Eqs.~(\ref{eq:rot_local})--(\ref{eq:pol_local}), for the four lowest-energy fully-relaxed DWs, belonging (in order of increasing energy) to the 109° easy, 71° hard, 71° easy, and 180° easy DW classes. The order parameters are displayed in a 25 Å window along the wall-normal coordinate $\xi$ surrounding a single DW. The edges of the plot window correspond approximately to the domain centers; thus, the plotted window spans roughly half of the DW supercell. Also, in Fig.~\ref{fig:dw_vesta_diag} we plot corresponding VESTA renderings of the corresponding 25 Å segments of our DW supercells, and mark the values of $\xi$ at which the polarization and octahedral rotation switch. In particular these diagrams permit comparison of the short vs. gradual change of the octahedral rotation pattern across the 109° easy and 71° hard walls.

First, we examine the 109° easy wall. Notably, the components of the polarization and rotation order parameters (which we also denote $P_\alpha(\xi)$ and $\phi_\alpha(\xi)$) that change sign do so at two different wall-normal coordinates $\xi$. The wall center (averaged between the switch coordinates of the two OPs) remains at the NbO$_2$ plane at which it was initialized, but the switch coordinate of the polarization and rotation OPs each migrate individually $\pm \frac{a}{2}$ away from the wall to opposite and adjacent NaO planes. Further, observe the thin nature of this DW: the polarization switches completely between the bulk $R3c$ states over a length of $<10$ Å, while the rotation pattern switches perfectly from one layer of octahedra to the next. Lubk postulated that the low-energy nature of the 109° easy wall is likely owed to the maximal flexibility afforded by the location of the wall at the apices of two neighboring layers of octahedra, as compared to the 71° and 180° walls, which are oriented parallel to octahedron edges \cite{lubk_first-principles_2009}. As suggested by the observations of Ref.~\cite{kp_electron_2025}, and replicated in our own first-principles investigations, the energetic difference between in-phase and out-of-phase octahedral rotations ($M_3^+$ vs $R_4^+$ rotations) is generally quite small in NNO. Evidently more gradual adjustments to the octahedral rotations are required in the other DW configurations. Also apparent in the OP profile of the wall are the equalities $P_x(\xi) = P_y(\xi)$ and $\phi_x(\xi) = \phi_y(\xi)$, enforced by the layer group symmetry.

\begin{figure*}[t!]
    \centering
    \includegraphics[width=\textwidth]{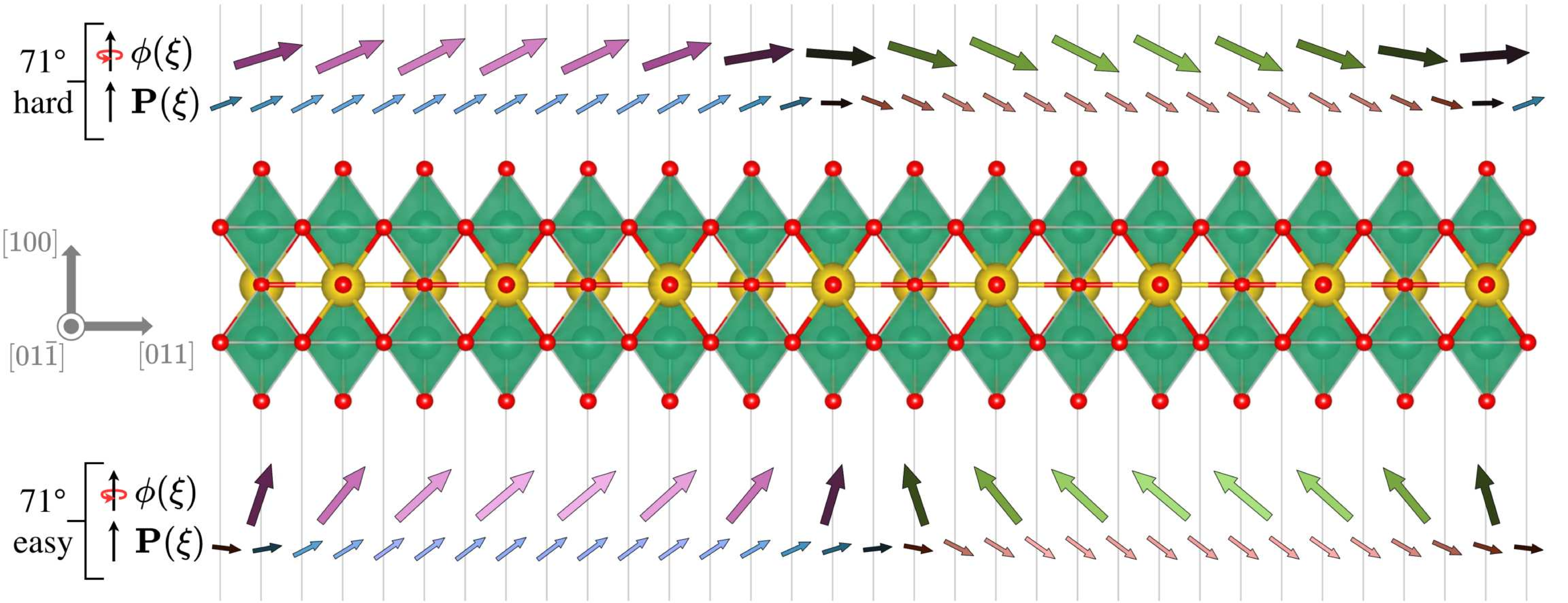}
    \caption{The orientation and magnitude of the axial-vector rotation order parameter $\bm{\phi}(\xi)$ and polar order parameter $\mathbf{P}(\xi)$ for the 71° hard (top row) and 71° easy (bottom row) DWs. The entire DW supercell is depicted here, with DWs visible near the middle and edges of the figure. In the center we depict the corresponding cubic-parent lattice, looking down $[0\bar{1}1]$; that is, with orientation matching the OP vector fields depicted. Each atomic plane and each layer of octahedra along $[011]$ in the reference $Pm\bar{3}m$ structure correspond respectively to a single plotted $\bm{\phi}$ and $\mathbf{P}$ order-parameter vector, indicated with the gray guidelines. The pink-green (blue-red) color gradients indicate the progression of the OP vectors between domains, with saturated color indicating the OP aligns with the bulk domain.}
    \label{fig:71_rotation_arrows}
\end{figure*}

In both the 71° easy and hard walls, the centers of the OP-switching profiles for $\bm{\phi}(\xi)$ and $\mathbf{P}(\xi)$ again migrate away from the $p=0$ initialization on an NaNbO plane, with centers approximately moving to opposite neighboring O$_2$ planes. As these walls also have layer group \#12, standard setting $pb11$, we observe similar symmetry-enforced equalities $P_y(\xi) = P_z(\xi)$ and $\phi_y(\xi) = \phi_z(\xi)$.

In this 71° hard wall, we observe that the order parameters do not saturate to their bulk values in the domain centers. The component $\phi_x$ of the rotation order parameter in particular takes almost the whole 20 Å window to switch between its saturated values, so such a DW in bulk $R3c$ would be at least 20 Å thick. This $\phi_x$ component is also observed to relax fully to zero below a critical domain thickness. Despite the apparently delicate energetics of this component of the rotation, this DW is the second-lowest in energy, behind only the 109° easy wall.

For the 71° easy wall, the change in octahedral rotation is compatible with the wall orientation, but the switch in rotation is still wider than the maximally-compatible 109° easy wall. An interesting feature of this particular DW is the pronounced bump in the nominally constant-valued $\phi_x(\xi)$ component, centered on the wall; despite having equal value in the domains on either side of the wall, some local increase in the $\phi_x(\xi)$ octahedral rotation component in the vicinity of the wall evidently improves the local structural environment at the wall. The same sort of perturbation is observed in the profile $\bm{\phi}(\xi)$ of the 109° hard wall, not plotted above, in the nominally constant-valued profile of $\phi_z(\xi)$. 

The common feature of these two walls is that a symmetry-enforced pair of rotation-vector components $\phi_\alpha(\xi)=\phi_\beta(\xi)$ jointly switch across the wall, while a free third component $\phi_\gamma(\xi)$ does not switch across the wall. Evidently, this compensation in $\phi_\gamma(\xi)$ occurs independently of the easy or hard character of the change in octahedral rotation. We propose that because this third free component of the rotation order parameter corresponds to rotation about a single Cartesian pseudocubic axis, i.e., about an axis passing through the apices of the octahedra, such perturbations do not come at a steep inherent energetic cost, regardless of the orientation of this axis relative to the DW.

In Fig.~\ref{fig:71_rotation_arrows} we provide a more explicit geometric intuition for the different OP profiles of the two 71° walls. The layer-group symmetry confines both the $\mathbf{P}(\xi)$ vector and $\bm{\phi}(\xi)$ pseudovector to lie in a shared plane, and as such we can diagram nicely the evolution of these order parameters across the DW with a quiver plot. Observe that the pseudovector $\bm{\phi}(\xi)$ must rotate through the $\hat{\mathbf{n}}$ direction in the hard wall, but not in the easy wall. The OP profile $\bm{\phi}(\xi)$ in the 71° wall has a distinct Néel-like character, where the pseudovector maintains a nearly constant magnitude across the wall thanks to the compensating adjustment of the component $\phi_x$, but its orientation changes, rotating in the plane of the two bulk domains.

In the 180° easy wall, the components of $\bm{\phi}(\xi)$ remain very nearly constant-valued across the DW, nearly fixed to their bulk values. The single nontrivial symmetry of the $p\bar{1}$ layer group of this DW is $(\bar{1}|000)$, which enforces $P_\alpha(\xi) = - P_\alpha(-\xi)$. Note, however, the approximate but not numerically exact equality of the components of polarization $P_y(\xi) \approx P_z(\xi)$. This is not enforced by the layer group symmetry, nor any spurious point symmetries from the layer group of the periodic DW supercell. Rather, $P_y(\xi) \approx P_z(\xi)$ implies that at coordinate $\xi$ there is approximately zero bound charge, 
$$\rho_b(\xi) = - \frac{d (\mathbf{P}\cdot\hat{\mathbf{n}})}{d \xi}$$
That the components $P_y\approx P_z$ do not saturate to the same mid-domain value as $P_x$ is indicative of finite-size-related effects. We also emphasize that while a local position-dependent polarization density is ill-defined in the modern theory, that $\mathbf{P}(\xi)$ produces zero bound charge everywhere in the DW supercell may indicate a real correspondence of this local polar order parameter to some aspect of the local crystal electrostatics.

Having characterized the local order-parameter textures, we now ask how the ferroelastic strain mismatch is accommodated across the DWs in our relaxed supercells.

\subsection{Effective inhomogeneous strain}

Recall our earlier mention of elastic compatibility. This condition ensures a planar interface between two strain states is possible, given the discontinuity between them at the wall. The DW supercell is not obviously capable of accommodating both such strain states by its own homogeneous strain, but there may be a component of the atomic displacement field away from the cubic parent which produces an equivalent local strain. For $R3c$, the change in strain state between a pair of domains across a wall corresponds to a change in sign of the equal off-diagonal rhombohedral strain components. The atomic displacement field to exactly reproduce this switch in the strain profile would look like a triangle wave, with period equal to the supercell length. 

In Fig.~\ref{fig:dw_local_strain} we plot the largest components of local strain, derived as described in Sec.~\ref{sec:analysis_methods}. Examining the plotted $\varepsilon_{yz}(\xi)=\varepsilon_{xz}(\xi)$ of the 109° easy wall, the expected switch between $\pm\varepsilon_{\text{bulk}}$ at the wall center (located at $\xi_\text{frac}=0$) is reproduced very closely. The long-range atomic displacement pattern that yields this strain profile is essentially a buckling in the $[110]_{\text{pc}}$ direction of the entire DW supercell. Taking the average value of $\varepsilon_{yz}=\varepsilon_{xz}$ for this wall over the initial plateau-like region in the plot, corresponding to the domain $[111]_+$, and adding this to the homogeneous infinitesimal strain tensor of the entire DW supercell with respect to the cubic parent, we obtain the approximate local strain state of the $[111]_+$ domain, as embedded in the 109° easy DW supercell,

$$ \varepsilon^{\text{DW}}_{[111]_+} = \begin{pmatrix} -9.66 & 5.96 & 5.63 \\ 5.96 & -9.66 & 5.63 \\ 5.63 & 5.63 & -9.52\end{pmatrix} \times 10^{-3}$$

This tensor lacks proper $R3c$ symmetry, but is quite close to the homogeneous infinitesimal strain tensor relating the fully-relaxed bulk $R3c$ lattice to the cubic parent,

$$ \varepsilon^{\text{bulk}}_{[111]_+} = \begin{pmatrix} -9.11 & 6.25 & 6.25 \\ 6.25 & -9.11 & 6.25 \\ 6.25 & 6.25 & -9.11\end{pmatrix} \times 10^{-3}$$

\section{DWs and $R3c \leftrightarrow Pbcm$ transition}
\label{sec:transition}

The two lowest-temperature bulk phases of NNO, $R3c$ and $Pbcm$, are notable for lacking a group-subgroup relationship. While not by any means uncommon in temperature-induced structural phase transitions, this is somewhat less typical of perovskites, and in particular the common alkali- and alkaline-earth-based families (Li,Na,K)(Nb,Ta)O$_3$ and (Ca,Sr,Ba)(Ti,Zr)O$_3$. As a result, the $Pbcm \rightarrow R3c$ structural phase transition does not have symmetry-prescribed candidate order parameters or sets of order parameters, whose condensation connects the two phases.

The complexity inherent in this transition has long appeared in crystallographic studies, showing hysteresis in the transitions $Pbcm \leftrightarrow R3c$ upon heating and cooling, as well as coexistence of these phases in a wide temperature range \cite{mishra_phase_2011,mishra_competing_2007,darlington_low-temperature_1973,jiang_local_2013}.

An early crystallographic work by Darlington and Megaw \cite{darlington_low-temperature_1973} contributes a powerful set of observations relating to the role of crystal twinning in this phase transition. They observed that an untwinned bulk $Pbcm$ phase will, upon lowering in temperature, consistently transition into a twinned $R3c$ phase, while a $Pbcm$ crystal which itself hosts significant twinning will tend to host an incomplete transition to a mixed $R3c$-and-$Pbcm$-type structure. These conclusions were supported by Jiang et al. in a more recent work \cite{jiang_local_2013}, who provided robust evidence from neutron scattering data of $R3c$ and $Pbcm$ coexistence in NNO, and suggested a locally $R3c$-like structural order with longer-range $Pbcm$-like order. Here, we show that the competing and coexisting $R3c$ and $Pbcm$-like order naturally admits interpretation in our language of domain walls, and use this picture to construct a plausible atomistic mechanism for the transition.

\begin{figure}[h!]
    \centering
    \includegraphics[width=\columnwidth]{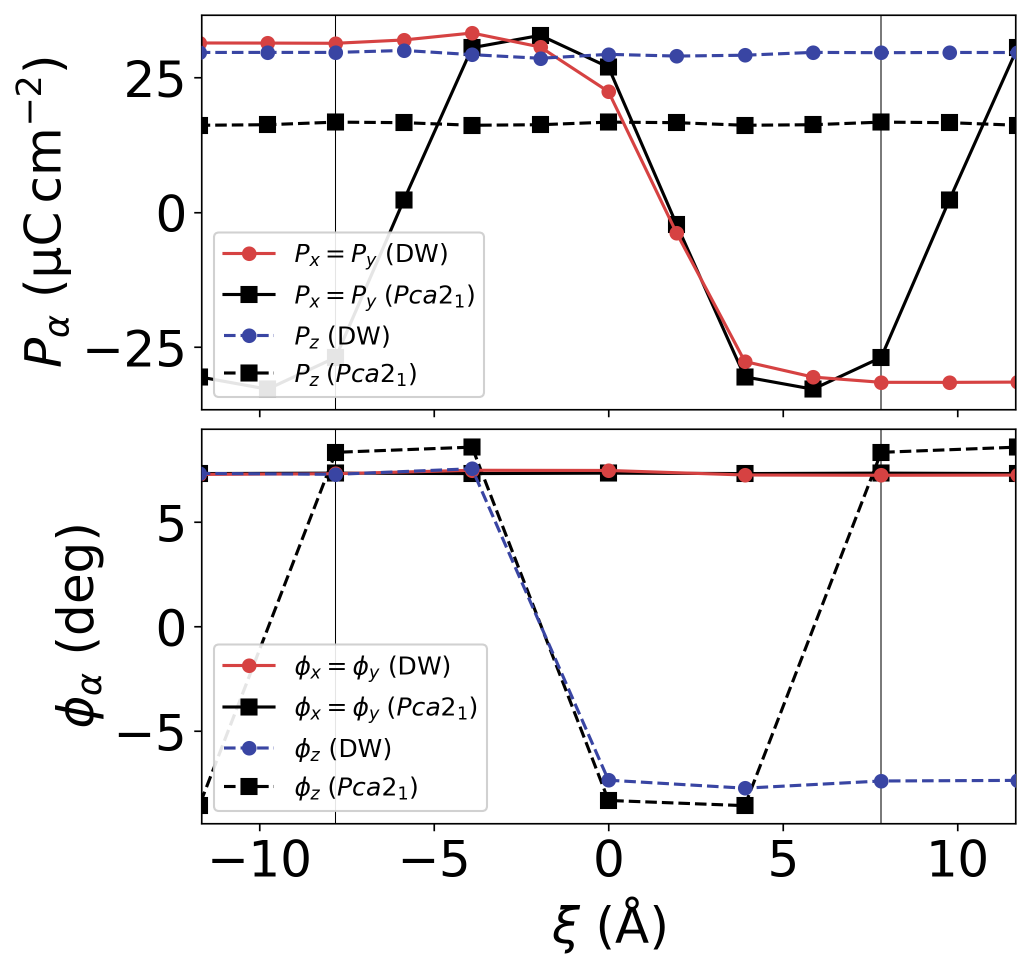}
    \caption{Projection of the atomic displacement pattern away from the $Pm\bar{3}m$ reference into the local OPs for polarization ($\mathbf{P}(\xi)$, top panel) and octahedral rotation ($\bm{\phi}(\xi)$, bottom panel) for the bulk $Pca2_1$ structure (black lines in both panels) and the 109° easy DW (colored lines). The faint black vertical lines indicate the boundaries of the $Pca2_1$ unit cell. $\xi$ here is the scalar coordinate along the DW normal $\hat{\mathbf{n}}$, equivalently along the c-axis direction of $Pca2_1$, which both point along the pseudocubic $[001]$ in this setting.}
    \label{fig:bulk_wall_OP_match}
\end{figure}

\begin{figure*}[t!]
    \centering
    \includegraphics[width=\textwidth]{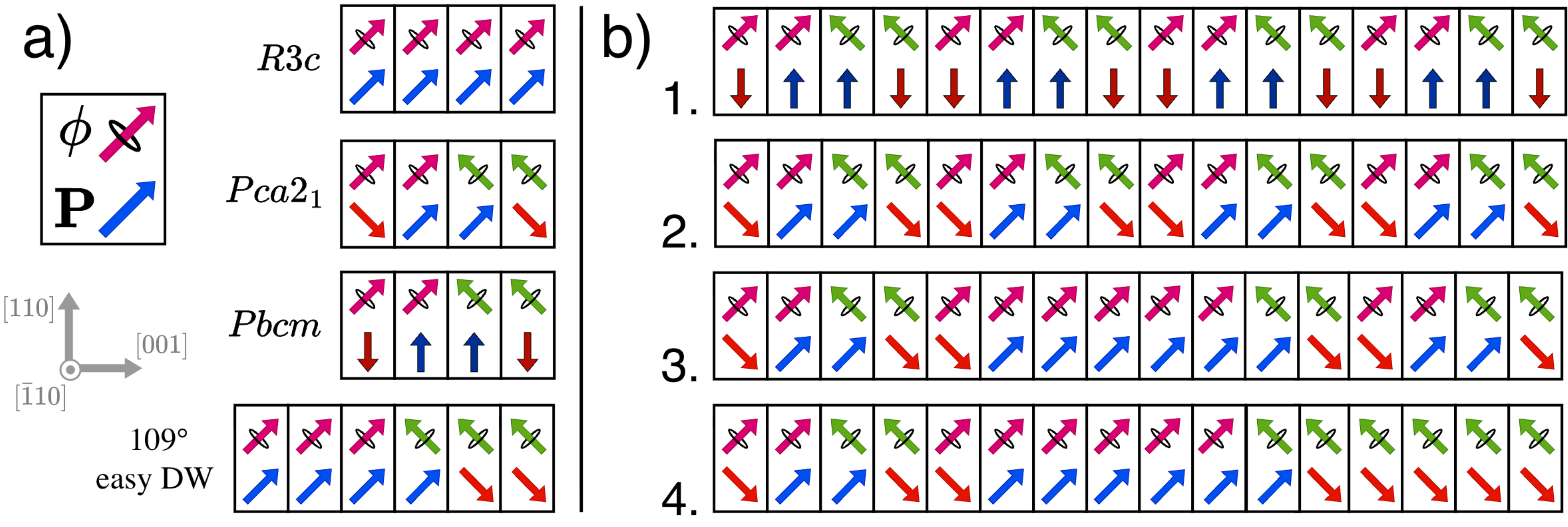}
    \caption{Local octahedral-rotation and (anti)polar cation patterns diagrammed for example structural phases. Each black rectangle represents a single $[001]$ oriented layer of 5-atom perovskite formula units. Top arrow and bottom arrow in the rectangle represent, respectively, the orientations of the $R_4^+$-like pseudovector rotation OP $\bm{\phi}$ and the polar OP $\mathbf{P}$. In (a), we diagram the local OP profiles for periodic cells of bulk $R3c$, $Pbcm$, and $Pca2_1$ structures, as well as the profile across a single 109° easy wall. In (b), we diagram the progression associated with our proposed structural transition mechanism $R3c \leftrightarrow Pbcm$ for a 16-layer-thick slice of bulk material, beginning from the top structure \textbf{\#1} which contains 4 layers of bulk $Pbcm$-structured NNO, down to the bottom \textbf{\#4}, a 109°-easy-twinned $R3c$ structure. Note that the OP arrow lengths are not to scale along the axes $[110]$ and $[001]$ in the plane of the figure. The arrows for OPs $\bm{\phi}$ and $\mathbf{P}$ have orientation-specific color coding of pink-green and blue-red respectively.}
    \label{fig:transition_mechanism}
\end{figure*}

\subsection{Compatibility between DWs and bulk structures}

Recall briefly the geometric motif of the room-temperature bulk $Pbcm$ structure and the related $Pca2_1$ polymorph: the octahedral rotation pattern, which we denoted $a^-a^-c^\pm$, corresponds to standard out-of-phase octahedral rotations of the $R_4^+$ mode about the $[100]$ and $[010]$ pseudocubic axes, and an alternating $++--$ type rotation about $[001]$, from the $T_4$ mode. The $\Delta_5$ mode consists of local polar displacements in the $[110]$ direction, also alternating like $++--$ while traveling along the $\mathbf{c}$ axis. The $Pca2_1$ polymorph features these motifs, plus an additional $\Gamma_4^-$ net polar displacement in the $[001]$ direction. We identify a geometric, symmetric, and energetic connection between these bulk orthorhombic structures and the 109° easy domain wall.

First, the thin 109° easy wall, across which we switch between the domains $[111]_+$ and $[11\bar{1}]_-$, hosts the same order parameter motifs as the bulk $Pca2_1$ structure, specifically the alternating $++--$ pattern in both the polarization components $P_x(\xi)=P_y(\xi)$ and octahedral rotation component $\phi_z(\xi)$. This correspondence is illustrated in Fig.~\ref{fig:bulk_wall_OP_match}, where we overlay the local OPs $\bm{\phi}(\xi)$ and $\mathbf{P}(\xi)$, as obtained from the $Pca2_1$ bulk structure and the 109° easy wall.

Second, it is possible to restore the full symmetry of the $Pca2_1$ space group by periodically stacking alternating $[111]_+$ and $[11\bar{1}]_-$ domains. These domains must be a) equal thickness, and b) contain an integer number of $R3c$ primitive units (that is, an even number of layers of oxygen octahedra). This can be derived by extending the order-parameter formalism in Sec.~\ref{sec:lg_theory}; in particular, extending the set of domain-exchanging symmetries as defined in Eq.~(\ref{eq:domain_exchange}) by permitting fractional translations along the now-translationally-symmetric axis $\mathbf{c} \parallel \hat{\mathbf{n}}$. Writing operations in reference to the setting of the domain wall supercell, this introduces new space group symmetries $(m_{010}|0\tfrac{1}{2}\tfrac{1}{2})$ and $(2_{001}|00\tfrac{1}{2})$, in addition to the symmetries $(1|000)$ and $(m_{100}|\tfrac{1}{2}\tfrac{1}{2}0)$ found in the layer group $pb11$ of the isolated 109° DW. This yields the space group $Pca2_1$, although the setting used to write the above operations is nonstandard and would correspond to the label $Pbc2_1$ of the same group.

The thin-domain limit of $[111]_+$ and $[11\bar{1}]_-$ periodic stacking is to stack domains two pseudocubic layers (two layers of octahedra) thick, along the $[001]$ axis. The primitive cell of such a domain stack is 40 atoms, with the aforementioned space group $Pca2_1$. In fact, initializing the primitive cell of this configuration using our domain-wall supercell method outlined in Sec.~\ref{sec:methods}, first-principles relaxation takes the structure exactly into the configuration of the bulk $Pca2_1$ room-temperature candidate polymorph.

Thus, in terms of structural geometry, symmetry, and energetics, the 4-cubic-cell-thick layer around an isolated 109° easy wall resembles a single layer of the bulk $Pca2_1$ orthorhombic structure, and in the thin-domain limit, the $Pca2_1$ structure is exactly recovered.

\subsection{Atomistic mechanism for $R3c \leftrightarrow Pbcm$ phase transition and coexistence}

With this compelling link between the 109° easy $R3c$ domain wall and the $Pca2_1$ bulk structure established, we proceed to discuss the phase transition $R3c \leftrightarrow Pbcm$. Darlington and Megaw \cite{darlington_low-temperature_1973} proposed that their X-ray diffraction photos of the low-temperature structure of NaNbO$_3$ corresponded to an $R3c$ structure, with equal fractions of two twin-variants related by a twofold rotation; the twinning they describe aligns exactly with the 109° domain wall, including the twin operation $2_{001}$ that connects our domains $[111]_+$ and $[11\bar{1}]_-$. This, and the established connection between these domain walls and the orthorhombic room-temperature structure, motivate us to propose an atomistic mechanism for the $R3c \leftrightarrow Pbcm$ phase transition. We will assume, for the sake of completeness, that the room-temperature structure is nonpolar, with $Pbcm$ order; the polar $Pca2_1$ subgroup of $Pbcm$ is essentially a waypoint of the transition pathway, and as such the process may proceed from either bulk polymorph. Note that we construct this pathway entirely from crystal symmetry and geometric arguments, and do not compute any energetic barriers associated with the transition or finite-temperature kinetics. 

We propose that the structural phase transition $Pbcm \rightarrow R3c$ ($R3c \rightarrow Pbcm$) can be modeled by a process equivalent to the annihilation (creation) of domain-wall pairs in 109°-easy-twinned $R3c$, in cooperation with the collective condensation (suppression) of net polarization in the $\hat{\mathbf{n}}$ domain-stacking direction of the system, here $[001]$. The schematic process for this is illustrated in Fig.~\ref{fig:transition_mechanism}, with a series of four related structures interpolating between structure \textbf{\#1}, pure bulk $Pbcm$ and structure \textbf{\#4}, bulk $R3c$ with multiple 109° easy domain walls. This figure depicts these structures by the orientation of the same local $R3c$ OPs $\bm{\phi}$ and $\mathbf{P}$ in each layer of a structure (one pseudocubic unit thick). We model this transition as, in the direction of lowering temperature, consisting of a sequential transition $Pbcm \rightarrow Pca2_1$ followed by a reorganization of local polarization+rotation units. The geometric process does not require a complete initial condensation of the polarization/complete transition to $Pca2_1$ from $Pbcm$, and this polarization could instead evolve cooperatively with the local reorganization of the octahedral rotations; we display this condensation of $[001]$ polarization as a first step for diagrammatic simplicity. Moving from the bulk $Pca2_1$-structured \textbf{\#2} to the structure \textbf{\#3}, the value $\bm{\phi}$ in two structural layers flips, and likewise the value $\mathbf{P}$ in a staggered sequence of two layers flips. The result is two domains of $R3c$, from the initial $Pca2_1$ which contained effectively four domains in the same volume; hence the interpretation of this model as ``domain wall pair annihilation.'' This process may continue, with reorganization of local $Pca2_1$-like units leading to the growth of $R3c$ structural domains, and perhaps the growth and annihilation of some of these larger domains as well. Barring any external energetic bias (electric field, strain, etc.) towards either one of the two $R3c$ domains which may nucleate here, a uniform bulk crystal of $Pbcm$-structured NNO could be expected to yield approximately equal fractions of each domain, exactly as observed by Darlington and Megaw. 

We can make a few further remarks to contextualize this scheme in the past crystallographic literature. Note that the observed coexistence of AFE $Pbcm$ and FE $Pmc2_1$ phases, if following the scheme proposed in previous literature \cite{htet_atomic_2022}, need not \textit{a priori} obstruct the transition to $R3c$, as the necessary local flips of $\bm{\phi}$ and condensation of nonzero components $P_z$ needed to produce an $R3c$ structure could still occur in $Pmc2_1$. However, we cannot comment on the height of the energetic barrier of this process, and whether it may differ between $Pbcm$, $Pmc2_1$ and the coexisting twinned phase of the two. In their work, Darlington and Megaw observed that a $Pbcm$ sample with extensive crystal twinning at room temperature produced, upon a decrease in temperature, a slow transition to a mixed $Pbcm$-$R3c$ phase, a mixed phase which has since been identified multiple times in the crystallographic literature \cite{jiang_local_2013,mishra_competing_2007}. Given our proposed mechanism consists essentially of local domain flips/domain creation and annihilation, this process is especially compatible with the observed hysteresis of the $Pbcm \leftrightarrow R3c$ transition upon heating and cooling \cite{mishra_competing_2007}. 

Given the mechanism we propose for this transition, or really any transition pathway with a spatially non-uniform character, the energetics of this phase transition will undoubtedly vary, depending on the character of the initial bulk room-temperature phase. Twinning in the room-temperature phase may produce elastic boundaries which impede this $R3c$ formation, as suggested in Ref.~\cite{darlington_low-temperature_1973}. More broadly, the very sensitive energetics of NaNbO$_3$ mean that twinning, crystal grain boundaries, sample morphology (powder particle size, bulk strain, etc.), and the concentration of vacancies or impurities may render some local crystal regions more or less amenable to undergoing the $Pbcm \rightarrow R3c$ transition as described. This aligns closely with the observation that the choice of synthesis method correlates with the level of $Pbcm$/$Pmc2_1$ coexistence in the room-temperature phase \cite{johnston_polar_2010}.

\section{Conclusion}

In this work, we undertook a systematic classification and study of the possible structural domain walls in the low-temperature $R3c$ phase of NaNbO$_3$. We introduced a symmetry-based classification of seven nontrivial domain wall classes, and found that four of these classes, those with 71° and 109° polarization angles, have a single sensible crystallographic orientation under the conditions of elastic compatibility and domain wall charge neutrality. First-principles calculations determined the 109° easy domain wall to have the lowest energy, only 9 mJ m$^{-2}$, comparable to the most favorable domain wall energies in simple ferroelectric BaTiO$_3$ \cite{meyer_ab_2002,zhang_first-principles_2023}. Furthermore, we find this 109° easy wall to be only around one cubic unit cell in thickness, while the other classes of domain walls may require thicknesses $>25$ Å. In the walls (71° easy and 109° hard) where two components $\phi_\alpha(\xi)=\phi_\beta(\xi)$ of the rotation order parameter switch between the domains, we find that the third component $\phi_\gamma(\xi)$, nominally constant-valued between the domains, develops a significant Néel-like compensating perturbation in the vicinity of the wall.

We observed that such a 109° easy domain wall is locally structured like the orthorhombic $Pca2_1$ polymorph, the polar variant of the common room-temperature $Pbcm$ structure, and used this structural compatibility to construct a sensible geometric pathway for the phase transition between room- and low-temperature phases $R3c \leftrightarrow Pbcm$, which lack a group-subgroup relationship and therefore lack a simple displacive-character transition. We argue that this phase transition mechanism is entirely compatible with past crystallographic results, namely that the low-temperature $R3c$ phase tends to consist of equal fractions of $2_{001}$-related crystal twins, that some NaNbO$_3$ samples transition incompletely to a mixed $R3c-Pbcm$ phase, and that the transition $R3c \leftrightarrow Pbcm$ is hysteretic upon heating and cooling \cite{darlington_low-temperature_1973,mishra_competing_2007,jiang_local_2013}. The work of Darlington and Megaw \cite{darlington_low-temperature_1973} remarkably observes or anticipates much of the character of this transition as we propose it here, while our results refine the microscopic mechanism of the transition. A natural extension of our discussion of the $R3c \leftrightarrow Pbcm$ phase transition would be the calculation of energy barriers along the proposed pathway, as well as treatment of this transition regime using finite-temperature methods.

In the narrowest sense, planar structural domain walls as described in this work can and do occur in NaNbO$_3$ and many other materials; crystal-structural ``twinning'' commonly occurs as two domains/twins stacked in alternating fashion along some crystallographic direction, corresponding in the domain wall picture to stacked planar domain walls of a single class. This is exactly the character of both the $Pbcm-Pmc2_1$ coexistence model \cite{htet_atomic_2022}, as well as the $Pbcm\rightarrow R3c$ phase transition as we construct it. More broadly, we advocate that such a study provides insight into the energetics and geometry associated with the coexistence of different types of local structural order, which may offer numerous insights beyond the case of well-defined crystal twinning.

\begin{acknowledgments}
J.G.N. and C.J.F. acknowledge support from the National Science Foundation Platform for the Accelerated Realization, Analysis, and Discovery of Interface Materials (PARADIM) under Cooperative Agreement No. DMR-2039380. C.J.F. supervised this project. J.G.N. carried out group-theoretic and first-principles calculations, and both authors analyzed these calculations. J.G.N. wrote this manuscript. This work was completed following the passing of C. J. Fennie. J.G.N. gratefully acknowledges his support, mentorship, and deep scientific insights, and dedicates this work to him.

\end{acknowledgments}

\appendix

\begin{table*}[t]
\caption{
Polar-vector irreducible representation $T_{1u}$ of the cubic point group
$O_h$, in the Cartesian basis $(x,y,z)$.
}
\label{table:T1u_Oh}
\centering

\begingroup
\scriptsize

\newcommand{\repM}[1]{%
  \begingroup
  \setlength{\arraycolsep}{3.2pt}%
  \renewcommand{\arraystretch}{1.05}%
  \begin{pmatrix}#1\end{pmatrix}%
  \endgroup
}

\setlength{\tabcolsep}{0pt}

\begin{tabular*}{\textwidth}
{@{\extracolsep{\fill}}cccccccc@{}}
\hline\hline

$1$ &
$2_{100}$ &
$2_{010}$ &
$2_{001}$ &
$3^{-}_{111}$ &
$3^{-}_{\bar{1}\bar{1}1}$ &
$3^{-}_{\bar{1}1\bar{1}}$ &
$3^{-}_{1\bar{1}\bar{1}}$
\\[-1pt]

$(x\,y\,z)$ &
$(x\,\bar y\,\bar z)$ &
$(\bar x\,y\,\bar z)$ &
$(\bar x\,\bar y\,z)$ &
$(y\,z\,x)$ &
$(y\,\bar z\,\bar x)$ &
$(\bar y\,\bar z\,x)$ &
$(\bar y\,z\,\bar x)$
\\[2pt]

$\repM{1&0&0\\0&1&0\\0&0&1}$ &
$\repM{1&0&0\\0&-1&0\\0&0&-1}$ &
$\repM{-1&0&0\\0&1&0\\0&0&-1}$ &
$\repM{-1&0&0\\0&-1&0\\0&0&1}$ &
$\repM{0&1&0\\0&0&1\\1&0&0}$ &
$\repM{0&1&0\\0&0&-1\\-1&0&0}$ &
$\repM{0&-1&0\\0&0&-1\\1&0&0}$ &
$\repM{0&-1&0\\0&0&1\\-1&0&0}$
\\[8pt]

$3^{+}_{111}$ &
$3^{+}_{\bar{1}1\bar{1}}$ &
$3^{+}_{\bar{1}\bar{1}1}$ &
$3^{+}_{1\bar{1}\bar{1}}$ &
$2_{1\bar{1}0}$ &
$4^{+}_{001}$ &
$4^{-}_{001}$ &
$2_{110}$
\\[-1pt]

$(z\,x\,y)$ &
$(z\,\bar x\,\bar y)$ &
$(\bar z\,x\,\bar y)$ &
$(\bar z\,\bar x\,y)$ &
$(\bar y\,\bar x\,\bar z)$ &
$(\bar y\,x\,z)$ &
$(y\,\bar x\,z)$ &
$(y\,x\,\bar z)$
\\[2pt]

$\repM{0&0&1\\1&0&0\\0&1&0}$ &
$\repM{0&0&1\\-1&0&0\\0&-1&0}$ &
$\repM{0&0&-1\\1&0&0\\0&-1&0}$ &
$\repM{0&0&-1\\-1&0&0\\0&1&0}$ &
$\repM{0&-1&0\\-1&0&0\\0&0&-1}$ &
$\repM{0&-1&0\\1&0&0\\0&0&1}$ &
$\repM{0&1&0\\-1&0&0\\0&0&1}$ &
$\repM{0&1&0\\1&0&0\\0&0&-1}$
\\[8pt]

$2_{01\bar{1}}$ &
$2_{011}$ &
$4^{+}_{100}$ &
$4^{-}_{100}$ &
$2_{\bar{1}01}$ &
$4^{-}_{010}$ &
$2_{101}$ &
$4^{+}_{010}$
\\[-1pt]

$(\bar x\,\bar z\,\bar y)$ &
$(\bar x\,z\,y)$ &
$(x\,\bar z\,y)$ &
$(x\,z\,\bar y)$ &
$(\bar z\,\bar y\,\bar x)$ &
$(\bar z\,y\,x)$ &
$(z\,\bar y\,x)$ &
$(z\,y\,\bar x)$
\\[2pt]

$\repM{-1&0&0\\0&0&-1\\0&-1&0}$ &
$\repM{-1&0&0\\0&0&1\\0&1&0}$ &
$\repM{1&0&0\\0&0&-1\\0&1&0}$ &
$\repM{1&0&0\\0&0&1\\0&-1&0}$ &
$\repM{0&0&-1\\0&-1&0\\-1&0&0}$ &
$\repM{0&0&-1\\0&1&0\\1&0&0}$ &
$\repM{0&0&1\\0&-1&0\\1&0&0}$ &
$\repM{0&0&1\\0&1&0\\-1&0&0}$
\\[8pt]

$\bar{1}$ &
$m_{100}$ &
$m_{010}$ &
$m_{001}$ &
$\bar{3}^{-}_{111}$ &
$\bar{3}^{-}_{\bar{1}\bar{1}1}$ &
$\bar{3}^{-}_{\bar{1}1\bar{1}}$ &
$\bar{3}^{-}_{1\bar{1}\bar{1}}$
\\[-1pt]

$(\bar x\,\bar y\,\bar z)$ &
$(\bar x\,y\,z)$ &
$(x\,\bar y\,z)$ &
$(x\,y\,\bar z)$ &
$(\bar y\,\bar z\,\bar x)$ &
$(\bar y\,z\,x)$ &
$(y\,z\,\bar x)$ &
$(y\,\bar z\,x)$
\\[2pt]

$\repM{-1&0&0\\0&-1&0\\0&0&-1}$ &
$\repM{-1&0&0\\0&1&0\\0&0&1}$ &
$\repM{1&0&0\\0&-1&0\\0&0&1}$ &
$\repM{1&0&0\\0&1&0\\0&0&-1}$ &
$\repM{0&-1&0\\0&0&-1\\-1&0&0}$ &
$\repM{0&-1&0\\0&0&1\\1&0&0}$ &
$\repM{0&1&0\\0&0&1\\-1&0&0}$ &
$\repM{0&1&0\\0&0&-1\\1&0&0}$
\\[8pt]

$\bar{3}^{+}_{111}$ &
$\bar{3}^{+}_{\bar{1}1\bar{1}}$ &
$\bar{3}^{+}_{\bar{1}\bar{1}1}$ &
$\bar{3}^{+}_{1\bar{1}\bar{1}}$ &
$m_{1\bar{1}0}$ &
$\bar{4}^{+}_{001}$ &
$\bar{4}^{-}_{001}$ &
$m_{110}$
\\[-1pt]

$(\bar z\,\bar x\,\bar y)$ &
$(\bar z\,x\,y)$ &
$(z\,\bar x\,y)$ &
$(z\,x\,\bar y)$ &
$(y\,x\,z)$ &
$(y\,\bar x\,\bar z)$ &
$(\bar y\,x\,\bar z)$ &
$(\bar y\,\bar x\,z)$
\\[2pt]

$\repM{0&0&-1\\-1&0&0\\0&-1&0}$ &
$\repM{0&0&-1\\1&0&0\\0&1&0}$ &
$\repM{0&0&1\\-1&0&0\\0&1&0}$ &
$\repM{0&0&1\\1&0&0\\0&-1&0}$ &
$\repM{0&1&0\\1&0&0\\0&0&1}$ &
$\repM{0&1&0\\-1&0&0\\0&0&-1}$ &
$\repM{0&-1&0\\1&0&0\\0&0&-1}$ &
$\repM{0&-1&0\\-1&0&0\\0&0&1}$
\\[8pt]

$m_{01\bar{1}}$ &
$m_{011}$ &
$\bar{4}^{+}_{100}$ &
$\bar{4}^{-}_{100}$ &
$m_{\bar{1}01}$ &
$\bar{4}^{-}_{010}$ &
$m_{101}$ &
$\bar{4}^{+}_{010}$
\\[-1pt]

$(x\,z\,y)$ &
$(x\,\bar z\,\bar y)$ &
$(\bar x\,z\,\bar y)$ &
$(\bar x\,\bar z\,y)$ &
$(z\,y\,x)$ &
$(z\,\bar y\,\bar x)$ &
$(\bar z\,y\,\bar x)$ &
$(\bar z\,\bar y\,x)$
\\[2pt]

$\repM{1&0&0\\0&0&1\\0&1&0}$ &
$\repM{1&0&0\\0&0&-1\\0&-1&0}$ &
$\repM{-1&0&0\\0&0&1\\0&-1&0}$ &
$\repM{-1&0&0\\0&0&-1\\0&1&0}$ &
$\repM{0&0&1\\0&1&0\\1&0&0}$ &
$\repM{0&0&1\\0&-1&0\\-1&0&0}$ &
$\repM{0&0&-1\\0&1&0\\-1&0&0}$ &
$\repM{0&0&-1\\0&-1&0\\1&0&0}$
\\

\hline\hline
\end{tabular*}

\endgroup
\end{table*}

\section{Matrix representations of $\Gamma_4^-$ and $R_4^+$}

\label{appendix:repres}
In Table~\ref{table:T1u_Oh} we present the 48 symmetry operations of the cubic point group $O_h$, including the label of each (im)proper rotation operation, the action on the coordinate triad $(x \,y\, z)$, and the matrix representatives $D^{T_{1u}}(R)$ of the $T_{1u}$ irreducible representation (the polar vector representation) of $O_h$, in its standard Cartesian form. These matrices may be used to construct the matrix representatives for representations $R_4^+$ and $\Gamma_4^-$ of the cubic space group $Pm\bar{3}m$ as:

\begin{subequations}
\label{eq:system}
\begin{align}
    D^{\Gamma_4^-}(R|\mathbf{t}) = D^{T_{1u}}(R) \label{eq:repGM} \\
    D^{R_4^+}(R|\mathbf{t}) = \text{det}(R)D^{T_{1u}}(R)e^{i \mathbf{k}_R \cdot \mathbf{t}} \label{eq:repR}
\end{align}
\end{subequations}

This expression makes clear the character of the two representations; $\Gamma_4^-$ describes transformations of a polar vector with trivial translations, while $R_4^+$ describes transformations of a pseudovector with translations modulated by wavevector $\mathbf{k}_R = \frac{2\pi}{a}(\tfrac{1}{2},\tfrac{1}{2},\tfrac{1}{2})$. All layer group calculations in this work were performed with 6-component order parameter vectors, transforming by the representation

$$D^{R_4^+ \oplus \Gamma_4^-}(R|\mathbf{t}) = D^{R_4^+ }(R|\mathbf{t}) \oplus D^{\Gamma_4^-}(R|\mathbf{t}) $$

 Note that while we use similar notation and procedure to that of Hatch and Stokes \cite{stokes_isotropy_nodate} for these representations, the bases of their representation matrices may generally be ordered differently from ours. We note also that formally the subspace of the atomic-displacement representation under which the full $R3c$ atomic distortion transforms contains four copies of the $\Gamma_4^-$ irrep, corresponding to different atomic-site-symmetry-resolved $\Gamma_4^-$ displacement modes. This implies that two different structures may both take an $R3c$ space group, but feature, e.g., different relative polar displacement of the A and B site sublattices. In initializing $R3c$ structures, we fix the ratios between each such site-specific mode amplitude to the values they take in the relaxed bulk $R3c$ structure, producing a single 3-component polar vector order parameter. In the numerical order parameter $\mathbf{P}(\xi)$, these atomic-site degrees of freedom are similarly integrated out, and the order parameter transforms as a single copy of $\Gamma_4^-$.

\section{Domain wall supercells}
\label{appendix:supercell}

Our discussion in Sec.~\ref{sec:lg_theory} indicates that for a given representative domain pair, there are multiple possible choices for the DW center $p$, which ostensibly represent multiple unique DW geometries, and may or may not result in a different wall configuration after first-principles structural relaxation. For the calculations in the main text, we exclusively initialize DW supercells with initial wall positions which we denote $p=0$.

For the 109° easy wall, the lowest-energy wall and the focus of our later analysis, we explicitly varied the initial value of $p$, and found two symmetry-inequivalent metastable wall configurations. The lower-energy configuration was initialized with our default choice, which we denoted $p=0$, with the relaxed wall having a composite center at this same $p=0$; recall, however, that the individual order parameters $\bm{\phi}$ and $\mathbf{P}$ switched at different positions $p_\phi = - \frac{a}{2}$ and $p_P = +\frac{a}{2}$. Any 109° wall initialized at $p \mod a = 0$ produces a symmetry-equivalent low-energy wall configuration, that is, any wall initialized at a $(001)$ NbO$_2$ plane. The higher energy metastable wall configuration tended to relax towards centers $p \mod a = \frac{a}{2}$, where instead of separate atomic planes, the order parameters $\bm{\phi}$ and $\mathbf{P}$ switch at the same NaO $(001)$ atomic plane. The energy of these other DWs is not significantly higher than the low-energy wall, at only 15 mJ m$^{-2}$.

Additionally, note that the DW supercells may possess additional point symmetries compared with the DW layer group, when supercells are initialized with two equal-thickness domains. Such symmetries generally consist of screws or glides that require $\tfrac{1}{2}$ fractional translation along the long $\mathbf{c}$ axis of the supercell; as a result, they do not considerably constrain the local crystal structure around a single DW, but rather relate the walls located at positions $0$ and $\tfrac{1}{2}$ to one another. We tested the relaxation of DWs with and without such spurious symmetries, and with and without the internal use of symmetry in the DFT code, and accordingly did not find appreciable differences in structure geometry or energetics.

\section{Local numerical order parameters}
\label{appendix:OP}

Here we detail the explicit construction of the position-dependent quantities $\mathbf{P}(\xi)$, $\bm{\phi}(\xi)$ and $\varepsilon_{ij}(\xi)$, implemented in our analysis of relaxed DW structures in the main text. Note that all numerical order parameters are derived by referencing the atomic positions of the cubic parent supercell with the positions of the relaxed DW supercell, with the domain-wall origin chosen such that there is zero mean atomic displacement from parent to child.

Define for each atomic layer $l$ with normal-coordinate $\xi_l$ a layer-summed Born electric charge (BEC) dipole, 

$$\mathbf{p}_l = e\sum_{\kappa \in l} \mathbf{Z}^*_\kappa \mathbf{u}_\kappa$$

Here, $\mathbf{p}_l$ is written in terms of the BEC tensors $\mathbf{Z}^*_\kappa$ of the cubic parent structure. Here we sum over the atoms in the $l$th layer along the supercell $\mathbf{c}$ axis defined according to the crystallographic planes of the parent $Pm\bar{3}m$ with the normal $\hat{\mathbf{n}}$. The local polar order parameter at an atomic plane along the DW normal is evaluated from this and the average of the BEC dipole of the two neighboring layers,

\begin{equation}
\label{eq:pol_local}
    \mathbf{P}(\xi_l) = \frac{1}{\Omega_l} \left[ \mathbf{p}_l + \frac{1}{2}(\mathbf{p}_{l-1}+\mathbf{p}_{l+1}) \right]
\end{equation}

This gives the local ``polarization'' averaged over an integer number of NaNbO$_3$ formula units in the volume $\Omega_l$ centered around the $l$th atomic plane. Note that $\mathbf{P}(\xi_l)$, while constructed as the linearized electric polarization due to one crystal layer, should not be conflated with a true bulk polarization in the sense of the modern theory \cite{resta_theory_2007}. That said, the domain-center local polarization as derived above should (and is in our work observed to) converge to a true bulk value, in the limit of large domains.

Similarly we obtain a layer-resolved profile of the octahedral rotations. For each octahedron in the cubic reference supercell, we match to the same octahedron in the DW supercell and apply a Kabsch rotation matrix algorithm to obtain the rotation vector $\boldsymbol{\theta}(\mathbf{R}_n)$ of the octahedron centered at cubic lattice vector $\mathbf{R}_n$. This rotation vector can be used to construct a local $R_4^+$-like order parameter amplitude, correcting by the alternating-sign phase factor from the wavevector $\mathbf{k}_R = \frac{2\pi}{a}(\frac{1}{2},\frac{1}{2},\frac{1}{2})$. Averaging this value over each plane of octahedra $l$ in the supercell, we obtain the local octahedral rotation 

\begin{equation}
    \label{eq:rot_local}
    \bm{\phi}(\xi_l) = \frac{1}{N_l} \sum_{n \in l} e^{-i\mathbf{k}_R \cdot \mathbf{R}_n} \bm{\theta}(\mathbf{R}_n)
\end{equation}

We find that this decomposition is well-conditioned, and that octahedra lying in the same plane do not tend to feature significantly different OP values $\bm{\phi}(\mathbf{R}_n)$. The octahedral rotation mode transforming as $R_4^+$ with amplitude $\bm{\phi}$ corresponds geometrically to the infinitesimal/linearized limit of the rotation described by the same rotation vector $\bm{\phi}$.

Finally, we are interested in the description of local (inhomogeneous) strain, which should manifest as a component of the atomic displacement field relating the cubic and DW supercells which varies smoothly and slowly in the wall-normal coordinate (with characteristic length scales much larger than the cubic lattice constant $a$), and is atomic-species independent. We produce a smooth, continuous atomic displacement field function by a Gaussian-kernel averaging procedure of all atomic displacements,

\begin{equation}
    \overline{\mathbf{u}}(\xi) = \frac{\sum_\kappa K_\sigma(\xi-\xi_\kappa) \mathbf{u}_\kappa}{\sum_\kappa K_\sigma(\xi-\xi_\kappa)}
\end{equation}

Here, $K_\sigma(\xi) = \exp(-\frac{\xi^2}{2\sigma^2})$. We find a value of 3 Å for $\sigma$ to produce the best result. From this, assuming any global rotation between the matched cubic and DW supercells has been eliminated, we obtain the local contribution to the infinitesimal strain tensor as:

\begin{equation}
    \label{eq:local_strain}
    \varepsilon_{ij}(\xi) = \frac{1}{2}\left( \frac{\partial \overline{u}_i}{\partial \xi}n_j + \frac{\partial \overline{u}_j}{\partial \xi}n_i\right)
\end{equation}

This expression depends on the scalar wall-normal coordinate $\xi$ (equivalently, the $\mathbf{c}$ axis coordinate of the DW supercell) and is written in terms of the components of the wall-normal vector $\hat{\mathbf{n}}$.

\bibliography{references}

@article{megaw_seven_1974,
	title = {The seven phases of sodium niobate},
	volume = {7},
	issn = {0015-0193, 1563-5112},
	url = {http://www.tandfonline.com/doi/abs/10.1080/00150197408237956},
	doi = {10.1080/00150197408237956},
	number = {1},
	urldate = {2026-08-10},
	journal = {Ferroelectrics},
	author = {Megaw, Helen D.},
	month = jan,
	year = {1974},
	pages = {87--89},
}

@article{darlington_low-temperature_1973,
	title = {The low-temperature phase transition of sodium niobate and the structure of the low-temperature phase, \textit{{N}}},
	volume = {29},
	issn = {0567-7408},
	url = {https://journals.iucr.org/paper?S0567740873006308},
	doi = {10.1107/S0567740873006308},
	number = {10},
	urldate = {2026-08-10},
	journal = {Acta Crystallographica Section B Structural Crystallography and Crystal Chemistry},
	author = {Darlington, C. N. W. and Megaw, H. D.},
	month = oct,
	year = {1973},
	pages = {2171--2185},
}

@article{glazer_studies_1973,
	title = {Studies of the lattice parameters and domains in the phase transitions of {NaNbO}$_{\textrm{3}}$},
	volume = {29},
	copyright = {http://journals.iucr.org/services/copyrightpolicy.html},
	issn = {0567-7394},
	url = {https://journals.iucr.org/paper?S0567739473001245},
	doi = {10.1107/S0567739473001245},
	number = {5},
	urldate = {2026-08-10},
	journal = {Acta Crystallographica Section A},
	author = {Glazer, A. M. and Megaw, H. D.},
	month = sep,
	year = {1973},
	pages = {489--495},
}

@article{sakowski-cowley_structure_1969,
	title = {The structure of sodium niobate at room temperature, and the problem of reliability in pseudosymmetric structures},
	volume = {25},
	issn = {0567-7408},
	url = {https://journals.iucr.org/paper?S0567740869003141},
	doi = {10.1107/S0567740869003141},
	number = {5},
	urldate = {2026-08-10},
	journal = {Acta Crystallographica Section B Structural Crystallography and Crystal Chemistry},
	author = {Sakowski-Cowley, A. C. and {\L}ukaszewicz, K. and Megaw, H. D.},
	month = may,
	year = {1969},
	pages = {851--865},
}

@article{lefkowitz_high-temperature_1966,
	title = {The high-temperature phases of sodium niobate and the nature of transitions in pseudosymmetric structures},
	volume = {20},
	copyright = {http://journals.iucr.org/services/copyrightpolicy.html},
	issn = {0365-110X},
	url = {https://journals.iucr.org/paper?S0365110X66001592},
	doi = {10.1107/S0365110X66001592},
	number = {5},
	urldate = {2026-08-10},
	journal = {Acta Crystallographica},
	author = {Lefkowitz, I. and {\L}ukaszewicz, K. and Megaw, H. D.},
	month = may,
	year = {1966},
	pages = {670--683},
}

@article{vousden_structure_1951,
	title = {The structure of ferroelectric sodium niobate at room temperature},
	volume = {4},
	copyright = {http://journals.iucr.org/services/copyrightpolicy.html},
	issn = {0365-110X},
	url = {https://journals.iucr.org/paper?S0365110X51001768},
	doi = {10.1107/S0365110X51001768},
	number = {6},
	urldate = {2026-08-10},
	journal = {Acta Crystallographica},
	author = {Vousden, P.},
	month = nov,
	year = {1951},
	pages = {545--551},
}

@article{campbell_isodisplace_2006,
	title = {\textit{{ISODISPLACE}} : a web-based tool for exploring structural distortions},
	volume = {39},
	issn = {0021-8898},
	shorttitle = {\textit{{ISODISPLACE}}},
	url = {https://journals.iucr.org/paper?S0021889806014075},
	doi = {10.1107/S0021889806014075},
	number = {4},
	urldate = {2026-06-25},
	journal = {Journal of Applied Crystallography},
	author = {Campbell, Branton J. and Stokes, Harold T. and Tanner, David E. and Hatch, Dorian M.},
	month = aug,
	year = {2006},
	pages = {607--614},
}

@article{fousek_orientation_1969,
	title = {The {Orientation} of {Domain} {Walls} in {Twinned} {Ferroelectric} {Crystals}},
	volume = {40},
	issn = {0021-8979, 1089-7550},
	url = {https://pubs.aip.org/jap/article/40/1/135/5465/The-Orientation-of-Domain-Walls-in-Twinned},
	doi = {10.1063/1.1657018},
	number = {1},
	urldate = {2026-06-23},
	journal = {Journal of Applied Physics},
	author = {Fousek, Jan and Janovec, V{\'a}clav},
	month = jan,
	year = {1969},
	pages = {135--142},
}

@misc{prandini_standard_2023,
	title = {A {Standard} {Solid} {State} {Pseudopotentials} ({SSSP}) library optimized for precision and efficiency},
	copyright = {Creative Commons Attribution 4.0 International},
	url = {https://archive.materialscloud.org/doi/10.24435/materialscloud:f3-ym},
	doi = {10.24435/MATERIALSCLOUD:F3-YM},
	urldate = {2026-06-11},
	publisher = {Materials Cloud},
	author = {Prandini, Gianluca and Marrazzo, Antimo and Castelli, Ivano E. and Mounet, Nicolas and Passaro, Elsa and Yu, Jusong and Marzari, Nicola},
	collaborator = {Prandini, Gianluca and Marzari, Nicola},
	month = apr,
	year = {2023},
	note = {Artwork Size: 35589539075
Pages: 35589539075},
}

@article{yang_understanding_2018,
	title = {Understanding and revisiting the most complex perovskite system via atomistic simulations},
	volume = {97},
	issn = {2469-9950, 2469-9969},
	url = {https://link.aps.org/doi/10.1103/PhysRevB.97.174106},
	doi = {10.1103/PhysRevB.97.174106},
	number = {17},
	urldate = {2026-06-11},
	journal = {Physical Review B},
	author = {Yang, Yali and Xu, Bin and Xu, Changsong and Ren, Wei and Bellaiche, Laurent},
	month = may,
	year = {2018},
	pages = {174106},
}

@article{jiang_local_2013,
	title = {Local structure of {NaNbO} 3 : {A} neutron scattering study},
	volume = {88},
	copyright = {http://link.aps.org/licenses/aps-default-license},
	issn = {1098-0121, 1550-235X},
	shorttitle = {Local structure of {NaNbO} 3},
	url = {https://link.aps.org/doi/10.1103/PhysRevB.88.014105},
	doi = {10.1103/PhysRevB.88.014105},
	number = {1},
	urldate = {2026-05-11},
	journal = {Physical Review B},
	author = {Jiang, Lu and Mitchell, D. C. and Dmowski, W. and Egami, T.},
	month = jul,
	year = {2013},
	pages = {014105},
}

@article{kp_electron_2025,
	title = {Electron ptychography reveals a ferroelectricity dominated by anion displacements},
	volume = {24},
	issn = {1476-1122, 1476-4660},
	url = {https://www.nature.com/articles/s41563-025-02205-x},
	doi = {10.1038/s41563-025-02205-x},
	number = {9},
	urldate = {2026-04-17},
	journal = {Nature Materials},
	author = {Kp, Harikrishnan and Xu, Ruijuan and Patel, Kinnary and Crust, Kevin J. and Khandelwal, Aarushi and Zhang, Chenyu and Prosandeev, Sergey and Zhou, Hua and Shao, Yu-Tsun and Bellaiche, Laurent and Hwang, Harold Y. and Muller, David A.},
	month = sep,
	year = {2025},
	pages = {1433--1440},
}

@article{lubk_first-principles_2009,
	title = {First-principles study of ferroelectric domain walls in multiferroic bismuth ferrite},
	volume = {80},
	copyright = {http://link.aps.org/licenses/aps-default-license},
	issn = {1098-0121, 1550-235X},
	url = {https://link.aps.org/doi/10.1103/PhysRevB.80.104110},
	doi = {10.1103/PhysRevB.80.104110},
	number = {10},
	urldate = {2026-04-17},
	journal = {Physical Review B},
	author = {Lubk, Axel and Gemming, S. and Spaldin, N. A.},
	month = sep,
	year = {2009},
	pages = {104110},
}

@article{htet_atomic_2022,
	title = {Atomic structural mechanism for ferroelectric-antiferroelectric transformation in perovskite {NaNbO} 3},
	volume = {105},
	issn = {2469-9950, 2469-9969},
	url = {https://link.aps.org/doi/10.1103/PhysRevB.105.174113},
	doi = {10.1103/PhysRevB.105.174113},
	number = {17},
	urldate = {2026-04-17},
	journal = {Physical Review B},
	author = {Htet, Cho Sandar and Nayak, S. and Manj{\'o}n-Sanz, Alicia and Liu, J. and Kong, J. and S{\o}rensen, D. R. and Marlton, F. and J{\o}rgensen, M. R. V. and Pramanick, A.},
	month = may,
	year = {2022},
	pages = {174113},
}

@article{johnston_polar_2010,
	title = {The {Polar} {Phase} of {NaNbO}$_{\textrm{3}}$ : {A} {Combined} {Study} by {Powder} {Diffraction}, {Solid}-{State} {NMR}, and {First}-{Principles} {Calculations}},
	volume = {132},
	issn = {0002-7863, 1520-5126},
	shorttitle = {The {Polar} {Phase} of {NaNbO}$_{\textrm{3}}$},
	url = {https://pubs.acs.org/doi/10.1021/ja101860r},
	doi = {10.1021/ja101860r},
	number = {25},
	urldate = {2026-04-17},
	journal = {Journal of the American Chemical Society},
	author = {Johnston, Karen E. and Tang, Chiu C. and Parker, Julia E. and Knight, Kevin S. and Lightfoot, Philip and Ashbrook, Sharon E.},
	month = jun,
	year = {2010},
	pages = {8732--8746},
}

@article{amisi_structural_2023,
	title = {Structural and dynamical phase transitions of {NaNbO} 3 from first-principles calculations},
	volume = {7},
	issn = {2475-9953},
	url = {https://link.aps.org/doi/10.1103/PhysRevMaterials.7.024408},
	doi = {10.1103/PhysRevMaterials.7.024408},
	number = {2},
	urldate = {2026-04-17},
	journal = {Physical Review Materials},
	author = {Amisi, Safari and Lambin, Philippe and Ghosez, Philippe},
	month = feb,
	year = {2023},
	pages = {024408},
}

@article{benedek_why_2013,
	title = {Why {Are} {There} {So} {Few} {Perovskite} {Ferroelectrics}?},
	volume = {117},
	issn = {1932-7447},
	url = {https://doi.org/10.1021/jp402046t},
	doi = {10.1021/jp402046t},
	number = {26},
	journal = {The Journal of Physical Chemistry C},
	publisher = {American Chemical Society},
	author = {Benedek, Nicole A. and Fennie, Craig J.},
	month = jul,
	year = {2013},
	pages = {13339--13349},
}

@article{hadaeghi_origin_2024,
	title = {Origin of antiferroelectricity in {NaNbO} 3},
	volume = {8},
	issn = {2475-9953},
	url = {https://link.aps.org/doi/10.1103/PhysRevMaterials.8.015004},
	doi = {10.1103/PhysRevMaterials.8.015004},
	number = {1},
	urldate = {2026-02-10},
	journal = {Physical Review Materials},
	author = {Hadaeghi, Niloofar and Dai, Mian and Zhang, Yixuan and Xie, Ruiwen and Nouri, Hamid and Zhang, Hongbin},
	month = jan,
	year = {2024},
	pages = {015004},
}

@article{ulinzheev_phase_1997,
	title = {Phase transitions induced in {NaNbO3} crystals by varying the direction of an external electric field},
	volume = {39},
	copyright = {http://www.springer.com/tdm},
	issn = {1063-7834, 1090-6460},
	url = {http://link.springer.com/10.1134/1.1129767},
	doi = {10.1134/1.1129767},
	number = {6},
	urldate = {2025-07-03},
	journal = {Physics of the Solid State},
	author = {Ulinzheev, A. V. and Leiderman, A. V. and Smotrakov, V. G. and Topolov, V. Yu. and Fesenko, O. E.},
	month = jun,
	year = {1997},
	pages = {972--974},
}

@article{zhang_revealing_2021,
	title = {Revealing the mechanism of electric-field-induced phase transition in antiferroelectric {NaNbO3} by \textit{in situ} high-energy x-ray diffraction},
	volume = {118},
	issn = {0003-6951, 1077-3118},
	url = {https://pubs.aip.org/apl/article/118/13/132903/40206/Revealing-the-mechanism-of-electric-field-induced},
	doi = {10.1063/5.0043050},
	number = {13},
	urldate = {2025-07-03},
	journal = {Applied Physics Letters},
	author = {Zhang, Mao-Hua and Zhao, Changhao and Fulanovi{\'c}, Lovro and R{\"o}del, J{\"u}rgen and Novak, Nikola and Sch{\"o}kel, Alexander and Koruza, Jurij},
	month = mar,
	year = {2021},
	pages = {132903},
}

@article{shiratori_polymorphism_2005,
	title = {Polymorphism in {Micro}-, {Submicro}-, and {Nanocrystalline} {NaNbO3}},
	volume = {109},
	issn = {1520-6106},
	url = {https://doi.org/10.1021/jp052974p},
	doi = {10.1021/jp052974p},
	number = {43},
	journal = {The Journal of Physical Chemistry B},
	publisher = {American Chemical Society},
	author = {Shiratori, Yosuke and Magrez, Arnaud and Dornseiffer, J{\"u}rgen and Haegel, Franz-Hubert and Pithan, Christian and Waser, Rainer},
	month = nov,
	year = {2005},
	pages = {20122--20130},
}

@incollection{resta_theory_2007,
	address = {Berlin, Heidelberg},
	title = {Theory of {Polarization}: {A} {Modern} {Approach}},
	volume = {105},
	isbn = {978-3-540-34590-9},
	shorttitle = {Theory of {Polarization}},
	url = {http://link.springer.com/10.1007/978-3-540-34591-6_2},
	doi = {10.1007/978-3-540-34591-6_2},
	urldate = {2025-04-25},
	booktitle = {Physics of {Ferroelectrics}},
	publisher = {Springer Berlin Heidelberg},
	author = {Resta, Raffaele and Vanderbilt, David},
	year = {2007},
	note = {Series Title: Topics in Applied Physics},
	pages = {31--68},
}

@article{zhang_first-principles_2023,
	title = {First-principles calculations of domain wall energies of prototypical ferroelectric perovskites},
	volume = {242},
	issn = {13596454},
	url = {https://linkinghub.elsevier.com/retrieve/pii/S1359645422007303},
	doi = {10.1016/j.actamat.2022.118351},
	urldate = {2025-02-23},
	journal = {Acta Materialia},
	author = {Zhang, Xueyou and Wang, Bo and Ji, Yanzhou and Xue, Fei and Wang, Yi and Chen, Long-Qing and Nan, Ce-Wen},
	month = jan,
	year = {2023},
	pages = {118351},
}

@article{schranz_polarization_2020,
	title = {Polarization of domain boundaries in {Sr} {Ti} {O} 3 studied by layer group and order-parameter symmetry},
	volume = {102},
	issn = {2469-9950, 2469-9969},
	url = {https://link.aps.org/doi/10.1103/PhysRevB.102.184101},
	doi = {10.1103/PhysRevB.102.184101},
	number = {18},
	urldate = {2025-02-08},
	journal = {Physical Review B},
	author = {Schranz, W. and Schuster, C. and Tr{\"o}ster, A. and Rychetsky, I.},
	month = nov,
	year = {2020},
	pages = {184101},
}

@article{mishra_phase_2011,
	title = {Phase stability and structural temperature dependence in sodium niobate: {A} high-resolution powder neutron diffraction study},
	volume = {83},
	copyright = {http://link.aps.org/licenses/aps-default-license},
	issn = {1098-0121, 1550-235X},
	shorttitle = {Phase stability and structural temperature dependence in sodium niobate},
	url = {https://link.aps.org/doi/10.1103/PhysRevB.83.134105},
	doi = {10.1103/PhysRevB.83.134105},
	number = {13},
	urldate = {2024-12-12},
	journal = {Physical Review B},
	author = {Mishra, S. K. and Mittal, R. and Pomjakushin, V. Yu. and Chaplot, S. L.},
	month = apr,
	year = {2011},
	pages = {134105},
}

@article{shiratori_temperature-induced_2007,
    title = {Temperature-induced {Phase} {Transitions} in {Micro}-, {Submicro}-, and {Nanocrystalline} {NaNbO}$_{\textrm{3}}$},
    volume = {111},
    issn = {1932-7447, 1932-7455},
    url = {https://pubs.acs.org/doi/10.1021/jp0738053},
    doi = {10.1021/jp0738053},
    number = {50},
    urldate = {2026-08-10},
    journal = {The Journal of Physical Chemistry C},
    author = {Shiratori, Yosuke and Magrez, Arnaud and Fischer, Werner and Pithan, Christian and Waser, Rainer},
    month = dec,
    year = {2007},
    pages = {18493--18502},
}

@article{ghanbari_strain-induced_2025,
    title = {Strain-induced lead-free morphotropic phase boundary},
    volume = {16},
    issn = {2041-1723},
    url = {https://www.nature.com/articles/s41467-025-63041-w},
    doi = {10.1038/s41467-025-63041-w},
    number = {1},
    urldate = {2026-08-10},
    journal = {Nature Communications},
    author = {Ghanbari, Reza and Kp, Harikrishnan and Patel, Kinnary and Zhou, Hua and Zhou, Tao and Liu, Rui and Wu, Liyan and Khandelwal, Aarushi and Crust, Kevin J. and Hazra, Sankalpa and Carroll, John and Meyers, Cedric J. G. and Wang, Jiayue and Prosandeev, Sergey and Qiao, Huimin and Kim, Young-Hoon and Nabei, Yoji and Chi, Miaofang and Sun, Dali and Balke, Nina and Holt, Martin and Gopalan, Venkatraman and Spanier, Jonathan E. and Muller, David A. and Bellaiche, Laurent and Hwang, Harold Y. and Xu, Ruijuan},
    month = aug,
    year = {2025},
    pages = {7766},
}

@article{hatch_complete_2001,
    title = {Complete listing of order parameters for a crystalline phase transition: {A} solution to the generalized inverse {Landau} problem},
    volume = {65},
    copyright = {http://link.aps.org/licenses/aps-default-license},
    issn = {0163-1829, 1095-3795},
    shorttitle = {Complete listing of order parameters for a crystalline phase transition},
    url = {https://link.aps.org/doi/10.1103/PhysRevB.65.014113},
    doi = {10.1103/PhysRevB.65.014113},
    number = {1},
    urldate = {2026-08-11},
    journal = {Physical Review B},
    author = {Hatch, Dorian M. and Stokes, Harold T.},
    month = dec,
    year = {2001},
    pages = {014113},
}

@article{meyer_ab_2002,
    title = {\textit{{Ab} initio} study of ferroelectric domain walls in {PbTiO} 3},
    volume = {65},
    copyright = {http://link.aps.org/licenses/aps-default-license},
    issn = {0163-1829, 1095-3795},
    url = {https://link.aps.org/doi/10.1103/PhysRevB.65.104111},
    doi = {10.1103/PhysRevB.65.104111},
    number = {10},
    urldate = {2026-08-25},
    journal = {Physical Review B},
    author = {Meyer, B. and Vanderbilt, David},
    month = mar,
    year = {2002},
    pages = {104111},
}

@article{mishra_competing_2007,
    title = {Competing antiferroelectric and ferroelectric interactions in {Na} {Nb} {O} 3 : {Neutron} diffraction and theoretical studies},
    volume = {76},
    copyright = {http://link.aps.org/licenses/aps-default-license},
    issn = {1098-0121, 1550-235X},
    shorttitle = {Competing antiferroelectric and ferroelectric interactions in {Na} {Nb} {O} 3},
    url = {https://link.aps.org/doi/10.1103/PhysRevB.76.024110},
    doi = {10.1103/PhysRevB.76.024110},
    number = {2},
    urldate = {2026-09-08},
    journal = {Physical Review B},
    author = {Mishra, S. K. and Choudhury, N. and Chaplot, S. L. and Krishna, P. S. R. and Mittal, R.},
    month = jul,
    year = {2007},
    pages = {024110},
}

@article{ahtee_structures_1972,
    title = {The structures of sodium niobate between 480{$^\circ$} and 575{$^\circ$}{C}, and their relevance to soft-phonon modes},
    volume = {26},
    issn = {0031-8086},
    url = {http://www.tandfonline.com/doi/abs/10.1080/14786437208226972},
    doi = {10.1080/14786437208226972},
    number = {4},
    urldate = {2026-09-14},
    journal = {Philosophical Magazine},
    author = {Ahtee, M. and Glazer, A. M. and Megaw, H. D.},
    month = oct,
    year = {1972},
    pages = {995--1014},
}

@article{glazer_structure_1972,
    title = {The structure of sodium niobate ({T}$_{\textrm{2}}$ ) at 600{$^\circ$}{C}, and the cubic-tetragonal transition in relation to soft-phonon modes},
    volume = {25},
    issn = {0031-8086},
    url = {http://www.tandfonline.com/doi/abs/10.1080/14786437208226856},
    doi = {10.1080/14786437208226856},
    number = {5},
    urldate = {2026-09-14},
    journal = {Philosophical Magazine},
    author = {Glazer, A. M. and Megaw, Helen D.},
    month = may,
    year = {1972},
    pages = {1119--1135},
}

@article{saito_lead-free_2004,
    title = {Lead-free piezoceramics},
    volume = {432},
    copyright = {http://www.springer.com/tdm},
    issn = {0028-0836, 1476-4687},
    url = {https://www.nature.com/articles/nature03028},
    doi = {10.1038/nature03028},
    number = {7013},
    urldate = {2026-09-14},
    journal = {Nature},
    author = {Saito, Yasuyoshi and Takao, Hisaaki and Tani, Toshihiko and Nonoyama, Tatsuhiko and Takatori, Kazumasa and Homma, Takahiko and Nagaya, Toshiatsu and Nakamura, Masaya},
    month = nov,
    year = {2004},
    pages = {84--87},
}

@article{zhu_ultrahigh_2025,
    title = {Ultrahigh piezoelectric performances of ({K},{Na}){NbO3} based ceramics enabled by structural flexibility and grain orientation},
    volume = {16},
    issn = {2041-1723},
    url = {https://www.nature.com/articles/s41467-025-56074-8},
    doi = {10.1038/s41467-025-56074-8},
    number = {1},
    urldate = {2026-09-15},
    journal = {Nature Communications},
    author = {Zhu, Li-Feng and Liu, Dong and Shi, Xiaoming and Deng, Shiqing and Liu, Jiecheng and Wei, Li-Yu and Yang, Zi-Qi and Wang, Qi and Zhang, Bo-Ping and Huang, Houbing and Zhang, Shujun and Li, Jing-Feng},
    month = jan,
    year = {2025},
    pages = {901},
}

@article{wang_large_2022,
    title = {Large piezoelectricity in {NaNbO}$_{\textrm{3}}$ -based lead-free ceramics \textit{via} tuning oxygen octahedral tilt},
    volume = {9},
    issn = {2051-6347, 2051-6355},
    url = {https://pubs.rsc.org/mh/article/9/3/1002-1009/732343},
    doi = {10.1039/D1MH01680F},
    number = {3},
    urldate = {2026-09-15},
    journal = {Materials Horizons},
    author = {Wang, Lu and Qi, He and Gao, Botao and Liu, Ye and Liu, Hui and Chen, Jun},
    year = {2022},
    pages = {1002--1009},
}

@article{qi_ultrahigh_2019,
    title = {Ultrahigh {Energy}-{Storage} {Density} in {NaNbO}$_{\textrm{3}}$ -{Based} {Lead}-{Free} {Relaxor} {Antiferroelectric} {Ceramics} with {Nanoscale} {Domains}},
    volume = {29},
    issn = {1616-301X, 1616-3028},
    url = {https://advanced.onlinelibrary.wiley.com/doi/10.1002/adfm.201903877},
    doi = {10.1002/adfm.201903877},
    number = {35},
    urldate = {2026-09-15},
    journal = {Advanced Functional Materials},
    author = {Qi, He and Zuo, Ruzhong and Xie, Aiwen and Tian, Ao and Fu, Jian and Zhang, Yi and Zhang, Shujun},
    month = aug,
    year = {2019},
    pages = {1903877},
}

@article{xie_superior_2026,
    title = {Superior {Capacitive} {Energy} {Storage} of {NaNbO}$_{\textrm{3}}$ -{Based} {MLCCs} {Enabled} by {Heterogeneous} {Short}-{Range} {Ferroic} {Orders}},
    volume = {36},
    issn = {1616-301X, 1616-3028},
    url = {https://advanced.onlinelibrary.wiley.com/doi/10.1002/adfm.202516297},
    doi = {10.1002/adfm.202516297},
    number = {8},
    urldate = {2026-09-15},
    journal = {Advanced Functional Materials},
    author = {Xie, Aiwen and Li, Zehao and Wu, Xiang and Tian, Ao and Zhang, Yi and Jiang, Xuewen and Xie, Xinchun and Gao, Xin and Er, Xiaokuo and Liu, Liqiang and Zhou, Cong and Zuo, Ruzhong},
    month = jan,
    year = {2026},
    pages = {e16297},
}

@article{jiang_boosting_2026,
    title = {Boosting energy storage in lead-free {NaNbO3}-based antiferroelectric ceramics through lamellar nanodomain engineering},
    volume = {12},
    issn = {23528478},
    url = {https://linkinghub.elsevier.com/retrieve/pii/S2352847825001443},
    doi = {10.1016/j.jmat.2025.101154},
    number = {2},
    urldate = {2026-09-15},
    journal = {Journal of Materiomics},
    author = {Jiang, Xuewen and Wang, Wei and Xie, Aiwen and Tang, Ziyi and Tian, Ao and Gao, Xin and Er, Xiaokuo and Liu, Liqiang and Zuo, Ruzhong},
    month = mar,
    year = {2026},
    pages = {101154},
}

@article{chen_phase_2022,
    title = {Phase engineering in {NaNbO3} antiferroelectrics for high energy storage density},
    volume = {8},
    issn = {23528478},
    url = {https://linkinghub.elsevier.com/retrieve/pii/S2352847822000375},
    doi = {10.1016/j.jmat.2022.03.004},
    number = {4},
    urldate = {2026-09-15},
    journal = {Journal of Materiomics},
    author = {Chen, Zhengu and Mao, Shuaifei and Ma, Li and Luo, Gengguang and Feng, Qin and Cen, Zhenyong and Toyohisa, Fujita and Peng, Xiuning and Liu, Laijun and Zhou, Huanfu and Hu, Changzheng and Luo, Nengneng},
    month = jul,
    year = {2022},
    pages = {753--762},
}

@article{cohen_origin_1992,
    title = {Origin of ferroelectricity in perovskite oxides},
    volume = {358},
    copyright = {http://www.springer.com/tdm},
    issn = {0028-0836, 1476-4687},
    url = {https://www.nature.com/articles/358136a0},
    doi = {10.1038/358136a0},
    number = {6382},
    urldate = {2026-09-15},
    journal = {Nature},
    author = {Cohen, Ronald E.},
    month = jul,
    year = {1992},
    pages = {136--138},
}

@article{shakhovoy_ferroelectric_2012,
    title = {Ferroelectric {Q} and antiferroelectric {P} phases' coexistence and local phase transitions in oxygen-deficient {NaNbO}$_{\textrm{3}}$ single crystal: micro-{Raman}, dielectric and dilatometric studies},
    volume = {43},
    copyright = {http://onlinelibrary.wiley.com/termsAndConditions\#vor},
    issn = {0377-0486, 1097-4555},
    shorttitle = {Ferroelectric {Q} and antiferroelectric {P} phases' coexistence and local phase transitions in oxygen-deficient {NaNbO}$_{\textrm{3}}$ single crystal},
    url = {https://analyticalsciencejournals.onlinelibrary.wiley.com/doi/10.1002/jrs.3140},
    doi = {10.1002/jrs.3140},
    number = {8},
    urldate = {2026-09-16},
    journal = {Journal of Raman Spectroscopy},
    author = {Shakhovoy, R. A. and Raevskaya, S. I. and Shakhovaya, L. A. and Suzdalev, D. V. and Raevski, I. P. and Yuzyuk, Yu. I. and Semenchev, A. F. and El Marssi, M.},
    month = aug,
    year = {2012},
    pages = {1141--1145},
}

@article{kim_explicating_2022,
    title = {Explicating the irreversible electric-field-assisted ferroelectric phase transition in the otherwise antiferroelectric sodium niobate for energy storage systems},
    volume = {10},
    issn = {2050-7526, 2050-7534},
    url = {https://pubs.rsc.org/tc/article/10/29/10500-10510/793095},
    doi = {10.1039/D2TC01817A},
    number = {29},
    urldate = {2026-09-16},
    journal = {Journal of Materials Chemistry C},
    author = {Kim, Kwangrae and Hwang, Woohyun and Lee, Ji-Hwan and Soon, Aloysius},
    year = {2022},
    pages = {10500--10510},
}

@article{zhang_electric-field-induced_2020,
    title = {Electric-field-induced antiferroelectric to ferroelectric phase transition in polycrystalline {NaNbO3}},
    volume = {200},
    issn = {13596454},
    url = {https://linkinghub.elsevier.com/retrieve/pii/S1359645420306960},
    doi = {10.1016/j.actamat.2020.09.002},
    urldate = {2026-09-16},
    journal = {Acta Materialia},
    author = {Zhang, Mao-Hua and Fulanovi{\'c}, Lovro and Egert, Sonja and Ding, Hui and Groszewicz, Pedro B. and Kleebe, Hans-Joachim and Molina-Luna, Leopoldo and Koruza, Jurij},
    month = nov,
    year = {2020},
    pages = {127--135},
}

@article{luo_well-defined_2023,
    title = {Well-defined double hysteresis loop in {NaNbO3} antiferroelectrics},
    volume = {14},
    issn = {2041-1723},
    url = {https://www.nature.com/articles/s41467-023-37469-x},
    doi = {10.1038/s41467-023-37469-x},
    number = {1},
    urldate = {2026-09-17},
    journal = {Nature Communications},
    author = {Luo, Nengneng and Ma, Li and Luo, Gengguang and Xu, Chao and Rao, Lixiang and Chen, Zhengu and Cen, Zhenyong and Feng, Qin and Chen, Xiyong and Toyohisa, Fujita and Zhu, Ye and Hong, Jiawang and Li, Jing-Feng and Zhang, Shujun},
    month = mar,
    year = {2023},
    pages = {1776},
}

@article{patel_properties_2021,
    title = {Properties of (001) {NaNbO} 3 films under epitaxial strain: {A} first-principles study},
    volume = {103},
    issn = {2469-9950, 2469-9969},
    shorttitle = {Properties of (001) {NaNbO} 3 films under epitaxial strain},
    url = {https://link.aps.org/doi/10.1103/PhysRevB.103.094103},
    doi = {10.1103/PhysRevB.103.094103},
    number = {9},
    urldate = {2026-09-17},
    journal = {Physical Review B},
    author = {Patel, Kinnary and Prosandeev, Sergey and Xu, Bin and Xu, Changsong and Bellaiche, L.},
    month = mar,
    year = {2021},
    pages = {094103},
}

@book{stokes_isotropy_1988,
    address = {Singapore New Jersey},
    title = {Isotropy subgroups of the 230 crystallographic space groups},
    isbn = {978-981-4434-31-7},
    publisher = {World Scientific},
    editor = {Stokes, Harold T. and Hatch, Dorian M.},
    year = {1988},
}

@misc{stokes_isotropy_nodate,
    title = {{ISOTROPY} {Software} {Suite}},
    url = {https://iso.byu.edu},
    author = {Stokes, H. T. and Hatch, D. M. and Campbell, B. J.},
}

@article{hatch_invariants_2003,
    title = {{INVARIANTS}: {Program} for {Obtaining} a {List} of {Invariant} {Polynomials} of the {Order} {Parameter} {Components} {Associated} with {Irreducible} {Representations} of a {Space} {Group}},
    volume = {36},
    doi = {10.1107/S0021889803005946},
    journal = {J. Appl. Crystallogr.},
    author = {Hatch, D. M. and Stokes, H. T.},
    year = {2003},
    pages = {951--952},
}

@article{aroyo_bilbao_I_2006,
	title = {Bilbao {Crystallographic} {Server}: {I}. {Databases} and crystallographic computing programs},
	volume = {221},
	issn = {2196-7105, 2194-4946},
	shorttitle = {Bilbao {Crystallographic} {Server}},
	url = {https://www.degruyterbrill.com/document/doi/10.1524/zkri.2006.221.1.15/html},
	doi = {10.1524/zkri.2006.221.1.15},
	number = {1},
	urldate = {2026-09-17},
	journal = {Zeitschrift f{\"u}r Kristallographie - Crystalline Materials},
	author = {Aroyo, Mois Ilia and Perez-Mato, Juan Manuel and Capillas, Cesar and Kroumova, Eli and Ivantchev, Svetoslav and Madariaga, Gotzon and Kirov, Asen and Wondratschek, Hans},
	month = jan,
	year = {2006},
	pages = {15--27},
}

@article{aroyo_bilbao_II_2006,
    title = {Bilbao {Crystallographic} {Server}. {II}. {Representations} of crystallographic point groups and space groups},
    volume = {62},
    issn = {0108-7673},
    url = {https://journals.iucr.org/paper?S0108767305040286},
    doi = {10.1107/S0108767305040286},
    number = {2},
    urldate = {2026-09-17},
    journal = {Acta Crystallographica Section A Foundations of Crystallography},
    author = {Aroyo, Mois I. and Kirov, Asen and Capillas, Cesar and Perez-Mato, J. M. and Wondratschek, Hans},
    month = mar,
    year = {2006},
    pages = {115--128},
}

@article{koruza_grain-size-induced_2017,
    title = {Grain-size-induced ferroelectricity in {NaNbO3}},
    volume = {126},
    issn = {13596454},
    url = {https://linkinghub.elsevier.com/retrieve/pii/S1359645416309879},
    doi = {10.1016/j.actamat.2016.12.049},
    urldate = {2026-09-21},
    journal = {Acta Materialia},
    author = {Koruza, Jurij and Groszewicz, Pedro and Breitzke, Hergen and Buntkowsky, Gerd and Rojac, Tadej and Mali{\v c}, Barbara},
    month = mar,
    year = {2017},
    pages = {77--85},
}

@article{wang_room-temperature_2026,
    title = {Room-temperature ferroelectricity in {NaNbO3} membrane},
    volume = {17},
    issn = {2041-1723},
    url = {https://www.nature.com/articles/s41467-026-74642-4},
    doi = {10.1038/s41467-026-74642-4},
    number = {1},
    urldate = {2026-09-21},
    journal = {Nature Communications},
    author = {Wang, Yunfan and Zheng, Zihao and Ren, Lele and Du, Yifeng and Guo, Jinming and Zhang, Shujun and Nan, Ce-Wen and Li, Bao-Wen},
    month = jun,
    year = {2026},
    pages = {7670},
}

@article{machado_relative_2011,
    title = {Relative phase stability and lattice dynamics of {NaNbO} 3 from first-principles calculations},
    volume = {84},
    copyright = {http://link.aps.org/licenses/aps-default-license},
    issn = {1098-0121, 1550-235X},
    url = {https://link.aps.org/doi/10.1103/PhysRevB.84.134107},
    doi = {10.1103/PhysRevB.84.134107},
    number = {13},
    urldate = {2026-09-21},
    journal = {Physical Review B},
    author = {Machado, R. and Sepliarsky, M. and Stachiotti, M. G.},
    month = oct,
    year = {2011},
    pages = {134107},
}

@article{padilla_first-principles_1996,
    title = {First-principles investigation of 180{$^\circ$} domain walls in {BaTi} {O} 3},
    volume = {53},
    copyright = {http://link.aps.org/licenses/aps-default-license},
    issn = {0163-1829, 1095-3795},
    url = {https://link.aps.org/doi/10.1103/PhysRevB.53.R5969},
    doi = {10.1103/PhysRevB.53.R5969},
    number = {10},
    urldate = {2026-09-21},
    journal = {Physical Review B},
    author = {Padilla, J. and Zhong, W. and Vanderbilt, David},
    month = mar,
    year = {1996},
    pages = {R5969--R5973},
}

@article{li_first-principles_2014,
    title = {First-principles study of 180 {$^\circ$} domain walls in {BaTiO} 3 : {Mixed} {Bloch}-{N{\'e}el}-{Ising} character},
    volume = {90},
    copyright = {http://link.aps.org/licenses/aps-default-license},
    issn = {1098-0121, 1550-235X},
    shorttitle = {First-principles study of 180 {$^\circ$} domain walls in {BaTiO} 3},
    url = {https://link.aps.org/doi/10.1103/PhysRevB.90.054106},
    doi = {10.1103/PhysRevB.90.054106},
    number = {5},
    urldate = {2026-08-26},
    journal = {Physical Review B},
    author = {Li, Menglei and Gu, Yijia and Wang, Yi and Chen, Long-Qing and Duan, Wenhui},
    month = aug,
    year = {2014},
    pages = {054106},
}

@article{kvasov_structure_2016,
    title = {Structure and pressure-induced ferroelectric phase transition in antiphase domain boundaries of strontium titanate from first principles},
    volume = {94},
    copyright = {http://link.aps.org/licenses/aps-default-license},
    issn = {2469-9950, 2469-9969},
    url = {https://link.aps.org/doi/10.1103/PhysRevB.94.054102},
    doi = {10.1103/PhysRevB.94.054102},
    number = {5},
    urldate = {2026-09-21},
    journal = {Physical Review B},
    author = {Kvasov, Alexander and Tagantsev, Alexander K. and Setter, Nava},
    month = aug,
    year = {2016},
    pages = {054102},
}

@article{janovec_group_1972,
    title = {Group analysis of domains and domain pairs},
    volume = {22},
    copyright = {http://www.springer.com/tdm},
    issn = {0011-4626, 1572-9486},
    url = {http://link.springer.com/10.1007/BF01690203},
    doi = {10.1007/BF01690203},
    number = {10},
    urldate = {2026-09-22},
    journal = {Czechoslovak Journal of Physics},
    author = {Janovec, V.},
    month = oct,
    year = {1972},
    pages = {974--994},
}

@article{wang_bifeo_2013,
  title = {{BiFeO} 3 {Domain} {Wall} {Energies} and {Structures}: {A} {Combined} {Experimental} and {Density} {Functional} {Theory} + {U} {Study}},
  volume = {110},
  copyright = {http://link.aps.org/licenses/aps-default-license},
  issn = {0031-9007, 1079-7114},
  shorttitle = {{BiFeO} 3 {Domain} {Wall} {Energies} and {Structures}},
  url = {https://link.aps.org/doi/10.1103/PhysRevLett.110.267601},
  doi = {10.1103/PhysRevLett.110.267601},
  number = {26},
  urldate = {2026-10-16},
  journal = {Physical Review Letters},
  author = {Wang, Yi and Nelson, Chris and Melville, Alexander and Winchester, Benjamin and Shang, Shunli and Liu, Zi-Kui and Schlom, Darrell G. and Pan, Xiaoqing and Chen, Long-Qing},
  month = jun,
  year = {2013},
  pages = {267601},
}

\end{document}